\documentclass[journal, twocolumn]{IEEEtran}

\usepackage{cite}
\usepackage{graphicx}
\usepackage{amsmath}
\usepackage{booktabs}
\usepackage{subfigure}

\usepackage[norelsize, linesnumbered, ruled, lined, boxed, commentsnumbered]{algorithm2e}      
\SetKwRepeat{Do}{do}{while}%
\SetKw{Kwin}{ in } 
\SetKw{KwTo}{ to }
\SetKw{KwGoto}{goto }

\usepackage{setspace}

\newcommand{\AlgRef}[1]{Alg. \ref{#1}}

\usepackage{amssymb}   
\usepackage{booktabs}
\usepackage{graphicx}
\usepackage{booktabs}
\usepackage{array}
\usepackage{adjustbox}
\usepackage{amsmath}
\usepackage{multirow}
\usepackage{tabularx}
\usepackage{tcolorbox}
\tcbuselibrary{breakable}
\usepackage[colorlinks=true,
linkcolor=blue,       
citecolor=blue,       
urlcolor=blue,        
filecolor=blue        
]{hyperref}
\begin{document}


\title{Single-Step Six-Dimensional Movable Antenna Reconfiguration for High-Mobility IoV: Modeling, Analysis, and Optimization}

\author{
    Maoxin Ji, Qiong Wu,~\IEEEmembership{Senior Member,~IEEE}, Pingyi Fan,~\IEEEmembership{Senior Member,~IEEE},\\ Kezhi Wang,~\IEEEmembership{Senior Member,~IEEE}, Wen Chen, ~\IEEEmembership{Senior Member,~IEEE}, Cui Zhang,\\ and Khaled B. Letaief, ~\IEEEmembership{Fellow,~IEEE}

    \thanks{
    	Part of this work has been submitted to IEEE International Conference on Communications (ICC), 24–28 May 2026, Glasgow, Scotland, UK.
    	
        Maoxin Ji and Qiong Wu are with the School of Internet of Things Engineering, Jiangnan University, Wuxi 214122, China (e-mail: maoxinji@stu.jiangnan.edu.cn, qiongwu@jiangnan.edu.cn).
        
        Pingyi Fan is with the Department of Electronic Engineering, State Key laboratory of Space Network and Communications, Beijing National Research Center for Information Science and Technology, Tsinghua University, Beijing 100084, China (e-mail: fpy@tsinghua.edu.cn).
        
        Kezhi Wang is with the Department of Computer Science, Brunel University, London, Middlesex UB8 3PH, U.K (e-mail: Kezhi.Wang@brunel.ac.uk).
        
        
        Wen Chen is with the Department of Electronic Engineering, Shanghai Jiao Tong University, Shanghai 200240, China (e-mail: wenchen@sjtu.edu.cn).
        
        Cui Zhang is with the School of Internet of Things Engineering, Wuxi Institute of Technology, wuxi, 214121, China (e-mail: zhangcui3@wxit.edu.cn).
        
        Khaled B. Letaief is with the Department of Electrical and Computer Engineering, the Hong Kong University of Science and Technology, Hong Kong (email: eekhaled@ust.hk).
        }
}
\maketitle
\begin{abstract}
	The Six-Dimensional Movable Antenna (6DMA) system has emerged as a promising technology to enhance wireless capacity by fully exploiting spatial degrees of freedom. However, applying 6DMA to high-mobility Internet of Vehicles (IoV) scenarios faces significant challenges, primarily due to the difficulty of acquiring instantaneous Channel State Information (CSI) and the risk of service interruptions caused by mechanical reconfiguration delays. To address these issues, this paper proposes a low-complexity, CSI-free single-step reconfiguration framework. First, we design a deterministic discrete position generation scheme based on a latitude-longitude grid with inherent topological structures. Leveraging graph theory, we explicitly model and theoretically derive the lower bounds of movement and time costs for antenna reconfiguration. Subsequently, utilizing the directional sparsity of 6DMA channels, we develop an adaptive optimization strategy that fuses offline environmental priors with online historical feedback. Furthermore, a periodic reconfiguration mechanism based on predicted cumulative vehicle distributions is introduced. By strictly restricting antenna adjustments to the first-order spatial neighborhood, the proposed single-step method effectively eliminates service interruptions. Simulation results demonstrate that the proposed scheme significantly outperforms traditional fixed and global-search-based benchmarks in terms of uplink sum rate, while incurring negligible mechanical overhead and latency, thereby validating its feasibility and robustness in highly dynamic vehicular networks.
\end{abstract}

\IEEEpeerreviewmaketitle

\section{Introduction}

\IEEEPARstart{T}{he} evolution of 6G communication technologies is significantly driving the transformation of the Internet of Vehicles (IoV) \cite{ref1, 001, 002, 003, 004, 005, 006, 007, 008}. IoV not only provides vehicle users with intelligent, convenient, and diversified travel experiences but also triggers an explosive growth in data transmission demands \cite{ref2, ref3, ref013, ref014, ref015, ref016, ref017, ref018, ref019, ref020}, posing severe challenges to the resource efficiency and deployment strategies of wireless networks \cite{10233705, 5720555}. To address this demand, Multiple-Input Multiple-Output (MIMO) \cite{ref4, ref5} and massive MIMO \cite{ref6, ref7} technologies have been successively introduced into the IoV domain. However, although increasing the number of antennas can significantly enhance array gain, the associated high hardware costs and radio frequency (RF) link power consumption restrict their large-scale application \cite{ref8, ref009, ref010, ref011, ref012}. Therefore, exploring spatial degrees of freedom to improve system efficiency under the premise of a limited antenna scale has become a critical issue to be resolved.

In recent years, Movable Antenna (MA) technology has received widespread attention as an emerging paradigm \cite{ref99}, \cite{ref9}, \cite{Li2025Joint}. Its core idea is to reconstruct the channel by adjusting antenna positions, thereby suppressing interference and enhancing signal power \cite{zhu2025multiuser}. Fluid Antenna \cite{ref10} and Two-Dimensional Movable Antenna (2DMA) \cite{ref11, li2025over} technologies have been proposed to enhance system performance by adjusting individual antenna positions to capture small-scale Channel State Information (CSI) variations. However, fluid antennas typically require rapid switching of individual antennas along a one-dimensional space to suppress interference, necessitating frequent position changes, which is difficult to apply in IoV networks with rapidly changing instantaneous channel information. While 2DMA technology allows antennas to move within a given two-dimensional plane, granting certain spatial degrees of freedom, it also faces the issue of frequent movement. In particular, using a single antenna as the movement unit results in high movement frequency and mechanical overhead.

Building on this, Shao \textit{et al.} proposed a novel Six-Dimensional Movable Antenna (6DMA) technology \cite{ref12}. In this architecture, all antennas are divided into multiple surfaces deploying rectangular antenna arrays, connected to a central processing unit (CPU) via extendable and rotatable rods, with flexible wires inside the rods for power supply and signal exchange \cite{shao20256dma}. Each surface can move in three-dimensional (3D) space, and its orientation can be rotated in 3D space, hence the name 6DMA. The 6DMA system possesses extremely high spatial flexibility. By adjusting the position and orientation of antenna surfaces to capture the spatial distribution of users, frequent adjustments are not required for scenarios with slowly varying spatial distributions \cite{ref13}. Moreover, moving rectangular antenna arrays as units rather than single antennas significantly reduces movement costs. 

Existing studies indicate that 6DMA significantly outperforms traditional Fixed Position Antennas (FPAs) and other mobile antenna schemes. In \cite{ref12}, it shows that in scenarios with clustered user distributions, the spectral efficiency of 6DMA is improved by approximately 497\% compared to FPA and 268\% compared to fluid antennas or 2DMA schemes. However, the ideal 6DMA supporting continuous spatial movement poses significant challenges in hardware implementation. To address this, in \cite{ref14}, it proposed a 6DMA model based on discrete positions and rotational states, utilizing a Fibonacci sphere method to generate a set of discrete positions satisfying physical constraints. Furthermore, in \cite{ref15}, it revealed the directional sparsity of 6DMA channels for the first time, i.e., an antenna surface with a specific configuration provides high gain only to users in specific spatial regions. This characteristic is determined jointly by the distribution of environmental scatterers and the antenna radiation pattern. In \cite{ref19}, it proposed a hierarchical movement scheme to reduce the reconfiguration cost of 6DMA. These works have laid a solid theoretical foundation for the development of 6DMA.

Despite the immense potential of 6DMA, its practical deployment faces severe challenges, particularly in the acquisition of CSI. Due to the flexible and variable antenna positions, traditional channel estimation methods are difficult to apply \cite{ref16}. Existing optimization schemes mostly rely on statistical CSI obtained via Monte Carlo sampling or heuristic methods that collect large amounts of measurement data \cite{ref17}. For instance, in \cite{ref12}, it estimated the average channel capacity via the Monte Carlo method and designed an alternating optimization scheme for antenna positioning. In \cite{ref18}, it designed a low-complexity 6DMA optimization algorithm based on statistical channel information. In \cite{ref14}, it adopted the Conditional Sample Mean (CSM) method to evaluate the benefit of 6DMA antenna positions without CSI and selected positions based on a greedy algorithm. However, this method requires sampling multiple antenna configurations to collect data under a fixed user distribution before updating, making it difficult to adapt to dynamic scenarios.

Applying 6DMA to IoV scenarios faces two primary difficulties. First, the high-speed mobility of vehicles causes drastic temporal variations in user distribution and instantaneous CSI, necessitating frequent antenna reconfiguration. Second, IoV services have strict requirements for high reliability and low latency, while the delay caused by mechanical antenna reconfiguration may lead to intermittent service interruptions. In addition, although existing reference has provided in-depth analyses of the spatial characteristics of 6DMA, there is a lack of explicit modeling and quantitative analysis regarding the feasibility of antenna movement, movement cost (energy consumption), and time cost (latency).

To address the aforementioned challenges, this paper proposes a low-complexity 6DMA single-step reconfiguration framework suitable for IoV\footnote{The source code is available at: \href{https://github.com/qiongwu86/Single-Step-6DMA-Reconfiguration-for-High-Mobility-IoV-Modeling-Analysis-and-Optimization}{https://github.com/qiongwu86/Single-Step-6DMA-Reconfiguration-for-High-Mobility-IoV-Modeling-Analysis-and-Optimization}.}. We first design a discrete position generation method with a natural topological structure and quantify the reconfiguration cost based on graph theory. Subsequently, utilizing prediction information of vehicle distribution, we propose a strategy to rapidly optimize antenna positions within the neighborhood, achieving continuous coverage of dynamic traffic flow with extremely low mechanical overhead. The main contributions of this paper are summarized as follows:

\begin{enumerate}
	\item We introduce 6DMA technology into high-dynamic IoV scenarios for the first time and establish a channel model based on multipath characteristics. A deterministic discrete position and rotation set construction scheme based on a latitude-longitude grid is proposed, which naturally possesses a clear neighbor topological relationship. Based on graph theory, utilizing Breadth-First Search (BFS) and a modified Hungarian algorithm, we define and theoretically derive the lower bounds of movement cost and time cost required for antenna configuration transitions. The analysis shows that, benefiting from the high connectivity of the discrete grid and the sparse distribution of antenna surfaces, these bounds are compact and achievable.
	
	\item To cope with the rapid fluctuations in user distribution caused by high-speed vehicle movement, an optimization perspective based on predicted cumulative distribution rather than instantaneous distribution is proposed. By extending the prediction time window, the cumulative user distribution captures the relatively stable distribution trends in the environment, thereby significantly reducing the frequency of antenna reconfiguration and the physical overhead of each reconfiguration, enhancing system robustness.
	
	\item Addressing the difficulty of acquiring instantaneous CSI and the high sampling overhead of traditional CSM methods, we propose an adaptive position optimization scheme fusing offline priors with online historical feedback. By exploiting the directional sparsity of 6DMA, an offline mapping library is established to link joint position-rotation states to effective coverage areas. During the online phase, historical service rates are used to dynamically correct mapping errors arising from environmental scattering changes, achieving efficient evaluation similar to CSM.
	
	\item To solve the problem of intermittent service interruptions caused by multi-step mechanical movements, the reconfiguration action space of antenna surfaces is strictly restricted to the first-order physical neighborhood. This mechanism ensures that each antenna surface requires at most one step of movement to complete the configuration in each decision cycle, fundamentally eliminating the risk of service interruption. Given that vehicle movement exhibits certain spatiotemporal regularities, it is feasible to substitute the low-frequency global reconfiguration strategy used in static environments with a high-frequency fine-tuning neighborhood movement strategy. Simulation results demonstrate the effectiveness of the proposed method.
\end{enumerate}

The rest of this paper is organized as follows.
Section~\ref{sec:system_model} presents the system model, detailing the discretized 6DMA architecture, physical constraints, and the channel model tailored for high-mobility IoV scenarios, and formulates a prediction-distribution-based periodic optimization problem under single-step constraints.
Section~\ref{sec:discrete_generation} establishes a deterministic position generation scheme based on a latitude–longitude grid and analyzes the reconfiguration cost using graph theory.
Section~\ref{method} proposes an adaptive position optimization framework that integrates offline environmental priors with online historical feedback to enable CSI-free decision-making.
Section~\ref{results} presents the simulation results and performance analysis.
Finally, Section~\ref{conclusion} concludes the paper.

\section{System Model}
\label{sec:system_model}

\begin{figure}[t]
	\centering
	\includegraphics[trim=0.5cm 0.5cm 0.5cm 0.5cm, clip, width=\columnwidth]{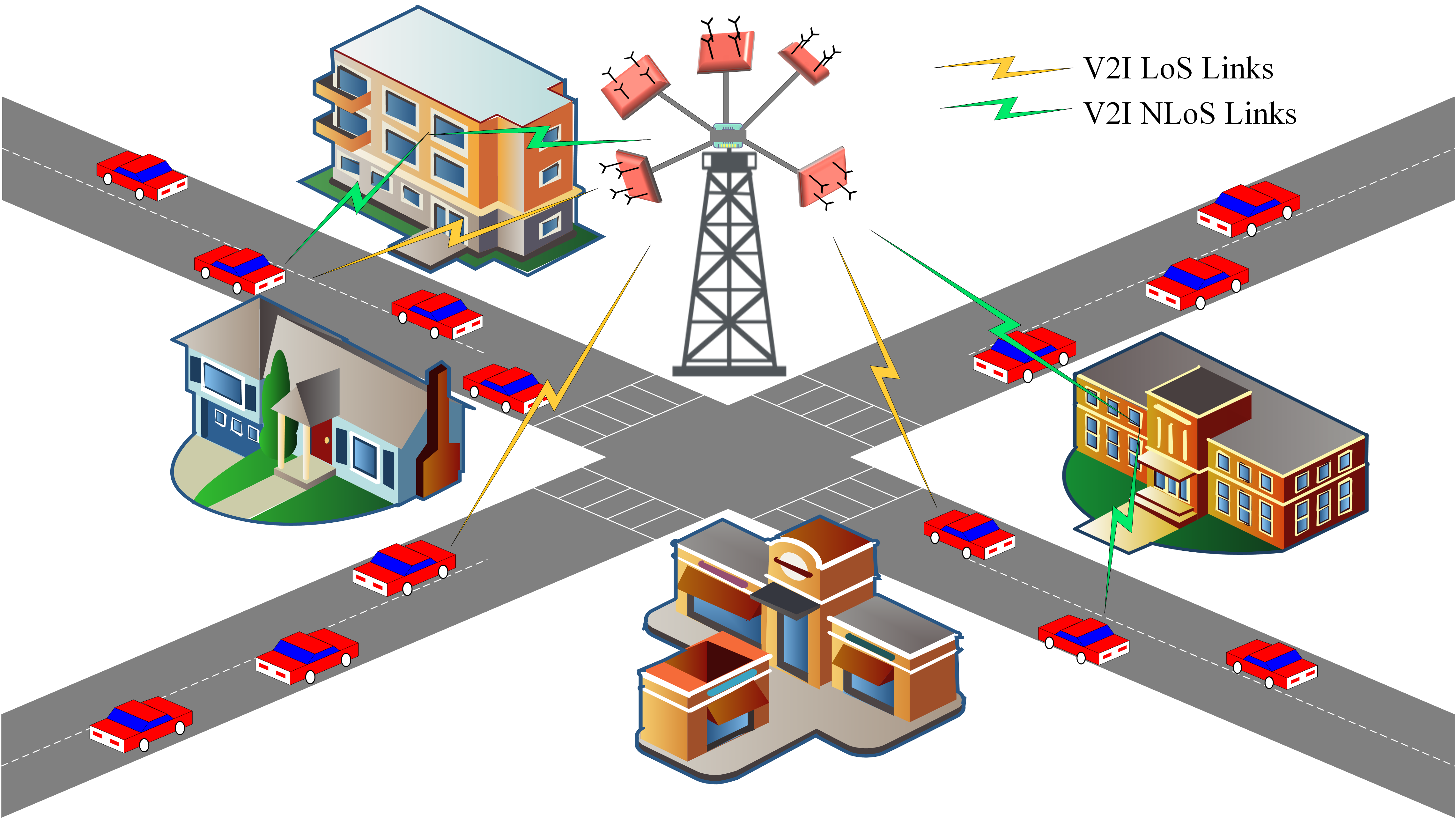}
	\caption{System Model}
	\label{system-model}
\end{figure}
This paper considers an uplink communication scenario at a typical urban intersection, as illustrated in Fig.~\ref{system-model}. Vehicle users are randomly distributed along two cross orthogonal roads and are served by a base station (BS) deployed at the center of the intersection. The network consists of a single BS and $K$ vehicles, with the set of vehicles denoted by $\mathcal{K} = \{1, 2, \ldots, K\}$. Each vehicle is equipped with a FPA, whereas the BS employs a receiving system based on 6DMAs.

The time domain is modeled using a discrete time-slot mechanism with a duration of $\Delta t$. We assume that the vehicle mobility exhibits quasi-static characteristics, i.e., the position $\mathbf{p}_k$ remains constant within a single time slot and is updated only at the slot boundaries based on velocities sampled from a truncated Gaussian distribution.
\vspace{-0.2cm}
\subsection{Discretized 6DMA Architecture and Constraints}

The 6DMA system comprises $U$ movable antenna surfaces connected to a CPU via extendable mechanical arms. Each surface is capable of independently adjusting its 3D spatial position and 3D rotational attitude. Constrained by mechanical precision and control complexity, the state space of the antenna surfaces is restricted to a discrete domain. Let $\mathcal{P} = \{(\mathbf{q}_i, \mathbf{r}_j) \mid i=1,\ldots,M; j=1,\ldots,J\}$ denote the complete set of all feasible configurations, where $M$ and $J$ represent the total numbers of discrete positions and rotational states, respectively. Specifically, $\mathbf{q}_i = [x_i, y_i, z_i]^\top \in \mathbb{R}^3$ denotes the center coordinates of the $i$-th candidate position, and $\mathbf{r}_j = [\alpha_j, \beta_j, \gamma_j]^\top$ is the Euler angle vector describing the rotation from the reference attitude (with the normal vector pointing towards $\mathbf{e}_x = [1,0,0]^\top$) to the current attitude. For any rotational state $j$, the corresponding unit outward normal vector $\mathbf{n}_j$ can be derived from the rotation matrix as:
\begin{equation}
	\mathbf{n}_j = \begin{bmatrix}
		\cos\beta_j \cos\gamma_j \\
		\sin\alpha_j \sin\beta_j \cos\gamma_j - \cos\alpha_j \sin\gamma_j \\
		\cos\alpha_j \sin\beta_j \cos\gamma_j + \sin\alpha_j \sin\gamma_j
	\end{bmatrix}.
\end{equation}

To describe the deployment state of the system, we introduce a binary decision matrix $\mathbf{Z} \in \{0,1\}^{M \times J}$. Specifically, $[\mathbf{Z}]_{i,j}=1$ indicates that an antenna surface is activated at position $i$ with rotational state $j$. The system activates a total of $U$ antenna surfaces (with $U \le M$), subject to the capacity constraint $\sum_{i=1}^M \sum_{j=1}^J [\mathbf{Z}]_{i,j} = U$. Furthermore, a physically realizable deployment scheme must simultaneously satisfy the following geometric and hardware constraints:

\begin{enumerate}
	\item Mutual Non-blocking Constraint: To prevent signal blockage or reflection interference between antenna surfaces, for any two activated surfaces $(i,j)$ and $(i',j')$ (i.e., $[\mathbf{Z}]_{i,j}=[\mathbf{Z}]_{i',j'}=1$ and $(i,j) \neq (i',j')$), the normal vector of surface $(i,j)$ must not point towards the location of surface $(i',j')$:
	\begin{equation}\label{11}
		\mathbf{n}_j^\top (\mathbf{q}_{i'} - \mathbf{q}_i) \leq 0.
	\end{equation}
	
	\item CPU Visibility Constraint: The radiation directions of all antenna surfaces must face away from the CPU holder located at the origin to avoid line-of-sight obstruction:
	\begin{equation}\label{22}
		\mathbf{n}_j^\top \mathbf{q}_i \geq 0, \quad \forall (i,j) \in \text{supp}(\mathbf{Z}).
	\end{equation}
	
	\item Minimum Separation Constraint: The Euclidean distance between any two activated positions must exceed a safety threshold $d_{\min}$ to prevent mechanical collision:
	\begin{equation}\label{3}
		\|\mathbf{q}_i - \mathbf{q}_{i'}\|_2 \geq d_{\min}, \quad \forall i \neq i'.
	\end{equation}
	
	\item Single Occupancy Constraint: At most one antenna surface can be deployed at each discrete position:
	\begin{equation}
		\sum_{j=1}^J [\mathbf{Z}]_{i,j} \leq 1, \quad \forall i=1,\ldots,M.
	\end{equation}
\end{enumerate}
\vspace{-0.4cm}
\subsection{Hybrid-Field Channel Modeling}

Consider the uplink transmission from vehicle $k$ to the BS. Each 6DMA surface integrates an FPA consisting of $Q$ elements, resulting in a total number of receiving elements $N_{\mathrm{total}} = U \times Q$. Given the large-aperture characteristic of the 6DMA, we adopt a hybrid near-field/far-field channel model \cite{ref20}. Specifically, signal propagation between surfaces is treated as far-field, while the phase differences among elements within a single surface retain near-field spherical wave characteristics.

First, a local coordinate system is established to describe the direction of arrival (DoA). Let $\mathbf{p}_k$ denote the coordinates of vehicle $k$. The unit line-of-sight (LoS) vector pointing towards antenna surface $(i,j)$ is given by $\hat{\mathbf{d}}_{k,i,j} = (\mathbf{p}_k - \mathbf{q}_i) / \|\mathbf{p}_k - \mathbf{q}_i\|$. Based on the reference axis $\mathbf{e}_{\text{ref}} = [0,0,1]^\top$, we construct the tangent plane orthogonal basis vectors $\mathbf{u}_{i,j} = (\mathbf{n}_j \times \mathbf{e}_{\text{ref}}) / \|\mathbf{n}_j \times \mathbf{e}_{\text{ref}}\|$ and $\mathbf{v}_{i,j} = \mathbf{n}_j \times \mathbf{u}_{i,j}$. The incident direction is projected onto the tangent plane as:
\begin{equation}
	\hat{\mathbf{d}}^{\parallel}_{k,i,j} = \frac{\hat{\mathbf{d}}_{k,i,j} - (\mathbf{n}_j^\top \hat{\mathbf{d}}_{k,i,j})\mathbf{n}_j}{\|\hat{\mathbf{d}}_{k,i,j} - (\mathbf{n}_j^\top \hat{\mathbf{d}}_{k,i,j})\mathbf{n}_j\|}.
\end{equation}
Consequently, the local elevation angle $\tilde{\theta}_{k,i,j}$ and azimuth angle $\tilde{\phi}_{k,i,j}$ are resolved as:
\begin{align}
	\tilde{\theta}_{k,i,j} &= \arccos(\mathbf{n}_j^\top \hat{\mathbf{d}}_{k,i,j}), \\
	\tilde{\phi}_{k,i,j} &= \operatorname{atan2}\left( \hat{\mathbf{d}}^{\parallel \top}_{k,i,j} \mathbf{v}_{i,j}, \hat{\mathbf{d}}^{\parallel \top}_{k,i,j} \mathbf{u}_{i,j} \right).
\end{align}

The antenna gain follows the 3GPP radiation pattern standard \cite{ref21}. The attenuation in the horizontal and vertical directions is defined as $A_H(\tilde{\phi}) = -\min(12(\tilde{\phi}/\phi_{3\mathrm{dB}})^2, A_m)$ and $A_V(\tilde{\theta}) = -\min(12(\tilde{\theta}/\theta_{3\mathrm{dB}})^2, A_m)$, respectively. The combined linear gain coefficient is expressed as:
\begin{equation}
	G_{k,i,j}^{(\mathrm{lin})} = 10^{(G_{\mathrm{max}} - \min(-(A_H + A_V), A_m))/10},
\end{equation}
where $\phi_{3\mathrm{dB}}$ and $\theta_{3\mathrm{dB}}$ represent the half-power beamwidths, $A_m$ denotes the sidelobe suppression ratio, and $G_{\max}$ is the peak gain.

The path loss model adopts the 3GPP UMi standard \cite{ref21}. For a path distance $d_{k,i} = \|\mathbf{p}_k - \mathbf{q}_i\|$, the LoS probability $p_{\mathrm{LoS}}(d_{k,i})$ is given by a piecewise exponential function (with parameters set to $d_1=18$ and $d_2=36$):
\begin{equation}
	p_{\mathrm{LoS}} (d_{k,i}) = \min\left(\frac{d_1}{d_{k,i}} ,1\right)(1-e^{-d_{k,i}/d_2 })+e^{-d_{k,i}/d_2 },
\end{equation}
where the propagation state is determined via Monte Carlo sampling. In the LoS state, the path loss $\text{PL}_{k,i,j,l}$ depends only on the distance and the carrier frequency $f_c$, which can be expressed as:
\begin{equation}
	\text{PL}_{k,i,j,l} = 32.4+21\log_{10} d_{k,i}+20\log_{10} f_c.
\end{equation}
In the non-LoS (NLoS) state, a log-normal shadow fading $\chi_l$ with a standard deviation of $\sigma_{SF}=7.82$ dB is superimposed, expressed as:
\begin{equation}
	\text{PL}_{k,i,j,l} = 35.3\log_{10} d_{k,i}+22.4+21.3\log_{10} f_c+\chi_l.
\end{equation}
The corresponding large-scale channel gain is $\eta_{k,i,j,l}^{(\mathrm{LS})} = 10^{-\text{PL}_{k,i,j,l}/10}$.

At the microscopic level, consider the $m$-th element on surface $(i,j)$, whose global coordinates are $\mathbf{c}_{i,j,m} = \mathbf{q}_i + u_m \mathbf{u}_{i,j} + v_m \mathbf{v}_{i,j}$, where $u_m$ and $v_m$ represent the relative displacements within the array. The near-field phase shift received by this element is $\phi_{k,i,j,m,l} = -\frac{2\pi}{\lambda} \|\mathbf{p}_k - \mathbf{c}_{i,j,m}\|$. Assuming the channel consists of a multipath set $\mathcal{D}_{k,i,j}$, where each path undergoes independent Rayleigh fading $\xi \sim \mathcal{CN}(0,1)$, the composite channel coefficient for this element is modeled as:
\begin{equation}
	h_{k,i,j,m} = \sum_{l \in \mathcal{D}_{k,i,j}} \sqrt{\eta_{k,i,j,l}^{(\mathrm{LS})} G_{k,i,j,l}^{(\mathrm{lin})}} \, e^{j\phi_{k,i,j,m,l}} \, \xi_{k,i,j,l}.
\end{equation}
This model accurately decouples the effects of array gain, large-scale fading, near-field phase, and small-scale fading.

To derive the aggregate channel vector, we first define the array response vector for a specific active surface $(i,j)$ satisfying $[\mathbf{Z}]_{i,j}=1$ as:
\begin{equation}
 	\mathbf{g}_{k,i,j} = [h_{k,i,j,1}, \ldots, h_{k,i,j,Q}]^\top \in \mathbb{C}^Q.
\end{equation}
By vertically concatenating the response vectors corresponding to all $U$ activated entries in $\mathbf{Z}$ (sorted by position index $i$ and rotation index $j$), the total channel vector $\mathbf{h}_k \in \mathbb{C}^{N_{\mathrm{total}}}$ for user $k$ is constructed. Consequently, the multi-user channel matrix is given by $\mathbf{H} = [\mathbf{h}_1, \ldots, \mathbf{h}_K]$. At this point, the channel matrix $\mathbf{H}(t)$ can be regarded as a nonlinear mapping function of the vehicle position distribution and the 6DMA configuration $\mathbf{Z}(t)$, concisely expressed as:
\begin{equation}
	\mathbf{H}(t) = \mathcal{H}(\{\mathbf{p}_k (t)\}_{k=1}^K, \mathbf{Z}(t)).
\end{equation}
Based on a matched filter receiver, the received signal-to-interference-plus-noise ratio (SINR) for user $k$ is given by:
\begin{equation} \label{SINR}
	\Gamma_k(\mathbf{H}) = \frac{P_k \|\mathbf{h}_k\|^2}{\sum_{n \neq k} P_n \frac{|\mathbf{h}_k^\mathrm{H} \mathbf{h}_n|^2}{\|\mathbf{h}_k\|^2 + \varepsilon} + \sigma^2},
\end{equation}
where $n \in \mathcal{K}$ represents the index of the interfering vehicle, and $\varepsilon$ is a regularization factor introduced for numerical stability. Finally, the total uplink sum rate of the system is calculated as $R_{\mathrm{sum}} = \sum_{k=1}^K B \log_2 (1 + \Gamma_k(\mathbf{H}))$.
\subsection{Dynamic Prediction and Optimization Problem Formulation}

Due to the high mobility of vehicles in IoV scenarios, the user spatial distribution exhibits rapid time-varying characteristics. Although the 6DMA system can adapt to such changes by flexibly adjusting the antenna configuration, real-time reconfiguration in every time slot is impractical due to the limited response speed of the mechanical servo system and computational resources. Therefore, this paper proposes a periodic single-step reconfiguration strategy based on motion prediction.

\subsubsection{Vehicle Kinematic Model}
Time is discretized into time slots of duration $\Delta t$. The position of vehicle $k$ in slot $t$, denoted by $\mathbf{p}_k(t) \in \mathbb{R}^3$, follows a linear kinematic model:
\begin{equation}
	\mathbf{p}_k(t+1) = \mathbf{p}_k(t) + v_k(t) \mathbf{a}_k(t) \Delta t,
\end{equation}
where $v_k(t) \geq 0$ is the instantaneous speed, and $\mathbf{a}_k(t) \in \mathbb{R}^3$ is the unit direction vector satisfying $\|\mathbf{a}_k(t)\| = 1$.

\subsubsection{Periodic Reconfiguration and State Prediction}
To balance system adaptability and implementation complexity, we set the antenna configuration $\mathbf{Z}$ to be updated every $N$ time slots. Define the set of time slots covered by the $l$-th reconfiguration period as $\mathcal{T}_l = \{t_l, \ldots, t_l + N - 1\}$, where the start time is $t_l \triangleq (l-1)N + 1$. During this period, the antenna configuration remains fixed, i.e., $\mathbf{Z}(t) = \mathbf{Z}_l, \forall t \in \mathcal{T}_l$.

Leveraging the spatiotemporal correlation of traffic flow, we employ a prediction mechanism based on the instantaneous average velocity $\bar{v}(t) = \frac{1}{K} \sum_{k=1}^K v_k(t)$ to estimate future trajectories. The predicted position $\hat{\mathbf{p}}_k$ is calculated recursively as:
\begin{equation}
	\hat{\mathbf{p}}_k(t+1) = \hat{\mathbf{p}}_k(t) + \mathbf{a}_k(t) \bar{v}(t) \Delta t,
\end{equation}
with the initial condition set to the current observation $\hat{\mathbf{p}}_k(t_l) = \mathbf{p}_k(t_l)$. Based on this predicted trajectory, the average sum rate for the $l$-th period is defined as:
\begin{equation}
	C_{\text{avg}}(\mathbf{Z}_l) \triangleq \frac{1}{N} \sum_{t \in \mathcal{T}_l} \sum_{k=1}^K B \log_2 \left( 1 + \Gamma_k \left( \mathbf{H}(\{\hat{\mathbf{p}}_k(t)\}, \mathbf{Z}_l) \right) \right).
\end{equation}

\subsubsection{Single-Step Reconfiguration Constraint and Problem Formalization} \label{question}
Although prediction-based reconfiguration improves long-term average performance, allowing antennas to jump arbitrarily within the global discrete space may result in significant mechanical displacement delays. Particularly in a discretized grid, cross-region adjustments often require multi-step mechanical actions, causing severe service interruptions. To address this, we propose a single-step reconfiguration mechanism. 

Let $\mathcal{I}_{l-1} = \{\text{pos}_1, \text{pos}_2, \dots, \text{pos}_U\}$ denote the set of position indices activated during the previous period, where $\text{pos}_u \in \{1, \dots, M\}$ represents the index of the spatial position occupied by the $u$-th antenna surface within the global discrete position set. To resolve the ambiguity of antenna correspondence and prevent infeasible long-distance swapping, we distinguish the $U$ antenna surfaces and track their individual trajectories. For the $u$-th antenna, its feasible search space for the current period is restricted to its \textit{closed neighborhood}:
\begin{equation}
	\bar{\mathcal{N}}_{\text{pos}_u} = \mathcal{N}_{\text{pos}_u} \cup \{\text{pos}_u\}, \quad u=1, \dots, U,
	\label{eq:local_space}
\end{equation}
where $\mathcal{N}_{\text{pos}_u}$ represents the set of physical neighbors of position $\text{pos}_u$ in the discrete grid. The definition of neighbors will be provided in a later section. This definition ensures that each antenna moves at most one step or remains stationary. 

Consequently, we formulate the optimization problem by introducing specific decision variables for each antenna to ensure their staying within their respective local neighborhoods. The problem is modeled as:
\begin{subequations}\label{opt:reconfig}
	\begin{align}
		\max_{\{\mathbf{Z}^{(u)}_l\}_{u=1}^U} \quad & C_{\text{avg}}\left(\sum_{u=1}^{U} \mathbf{Z}^{(u)}_l\right) \label{eq:prob_obj} \\
		\text{s.t.} \quad 
		& \sum_{j=1}^{J} [\mathbf{Z}^{(u)}_l]_{i,j} \in \{0, 1\}, \quad \forall i, u, \label{eq:prob_binary} \\
		& \sum_{i=1}^{M} \sum_{j=1}^{J} [\mathbf{Z}^{(u)}_l]_{i,j} = 1, \quad \forall u, \label{eq:prob_active} \\
		& [\mathbf{Z}^{(u)}_l]_{i,j} = 0, \quad \forall i \notin \bar{\mathcal{N}}_{\text{pos}_u}, \forall u, \label{eq:prob_local} \\
		& \sum_{u=1}^{U} \sum_{j=1}^{J} [\mathbf{Z}^{(u)}_l]_{i,j} \leq 1, \quad \forall i, \label{eq:prob_collision} \\
		& \text{Physical Constraints \eqref{11}, \eqref{22}, \eqref{3}}, \label{eq:prob_phys} 
	\end{align}
\end{subequations}
where $\mathbf{Z}^{(u)}_l \in \{0,1\}^{M \times J}$ represents the configuration matrix specifically for the $u$-th antenna. Constraint \eqref{eq:prob_active} ensures each antenna selects exactly one position-rotation state. Constraint \eqref{eq:prob_local} strictly limits the $u$-th antenna to its local neighborhood $\mathcal{S}_u$, effectively eliminating the risk of logical teleportation between disjoint regions. Constraint \eqref{eq:prob_collision} prevents collisions by ensuring that no two antennas occupy the same position, even if their neighborhoods overlap. Finally, the aggregate configuration $\mathbf{Z}_l = \sum_{u=1}^U \mathbf{Z}^{(u)}_l$ is used to evaluate the physical constraints and system capacity.

Although the above problem is formally closed, the precise acquisition of $\mathbf{H}$ requires enormous pilot overhead due to the complexity of 6DMA channels. Furthermore, the objective function exhibits high non-convexity and non-linearity, making direct numerical solutions infeasible in millisecond-level IoV scenarios. To tackle this challenge, in the following sections, we exploit the beam direction sparsity of 6DMA to propose a fast heuristic algorithm that does not require instantaneous CSI.
\vspace{-0.1cm}
\section{Discrete Position Generation and Cost Modeling for Movement and Time} \label{section3}
\label{sec:discrete_generation}
\subsection{Discrete Position Generation Based on Latitude-Longitude Grid}
\begin{figure*}[t]
	\centering
	\begin{minipage}[b]{0.28\textwidth}
			\centering
			\includegraphics[width=\textwidth]{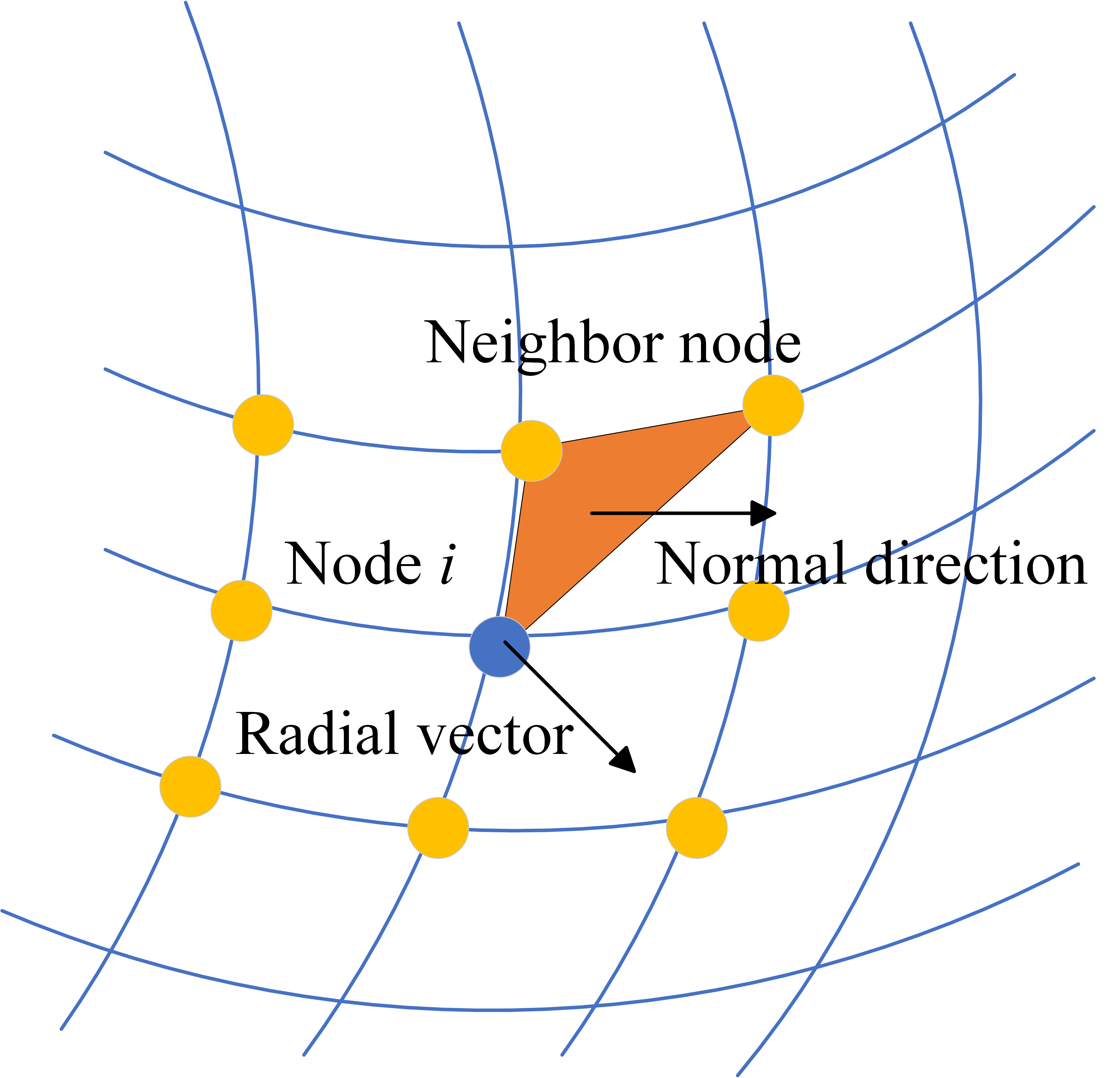}
			\caption{Illustration of the 8-neighbor topology and normal vector generation for general positions.}
			\label{fig9}
		\end{minipage}
	\hfill
	\begin{minipage}[b]{0.28\textwidth}
			\centering
			\includegraphics[width=\textwidth]{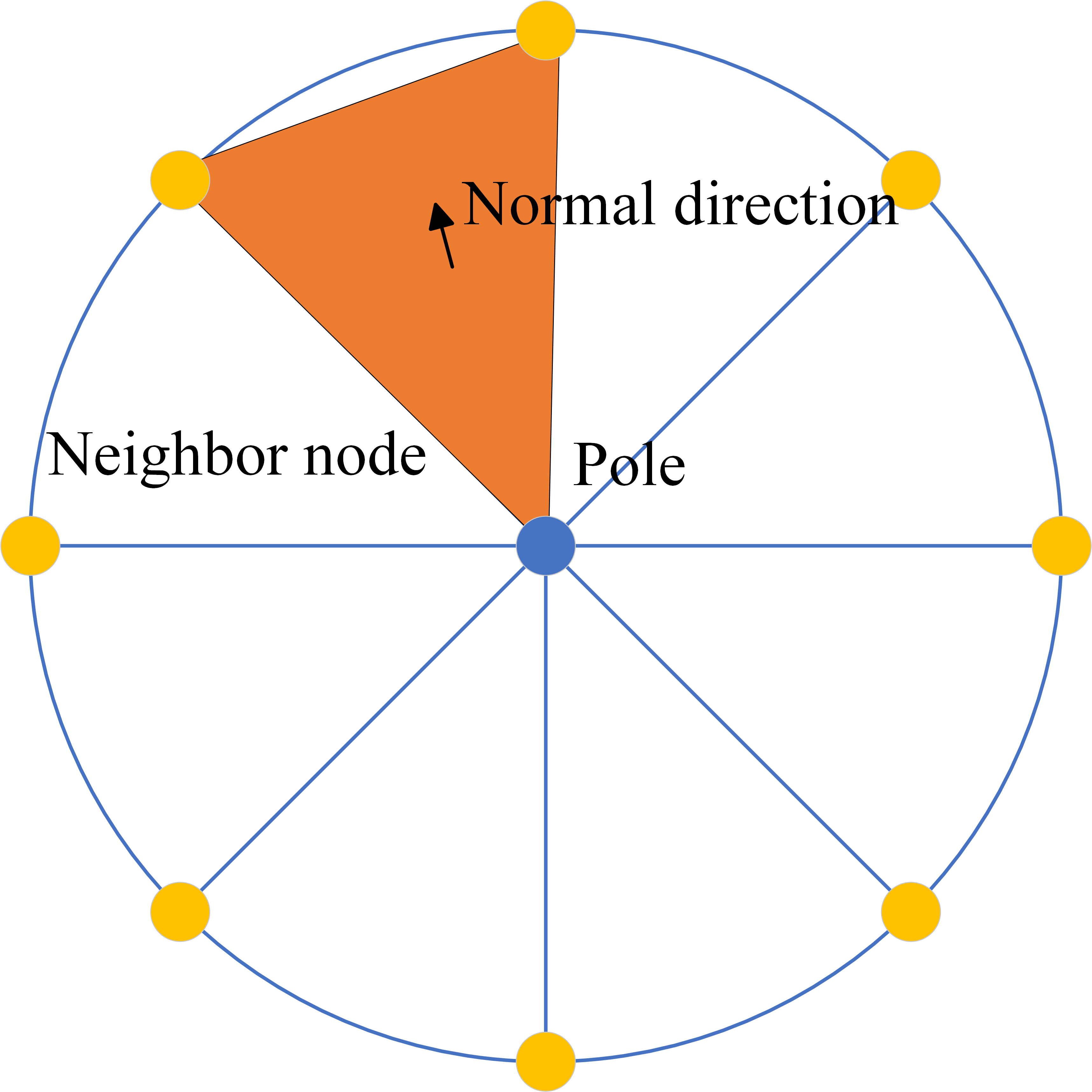}
			\caption{Neighbor definition and discrete orientation generation for the polar positions.}
			\label{fig10}
		\end{minipage}
	\hfill
	\begin{minipage}[b]{0.35\textwidth}
			\centering
			\includegraphics[width=\textwidth]{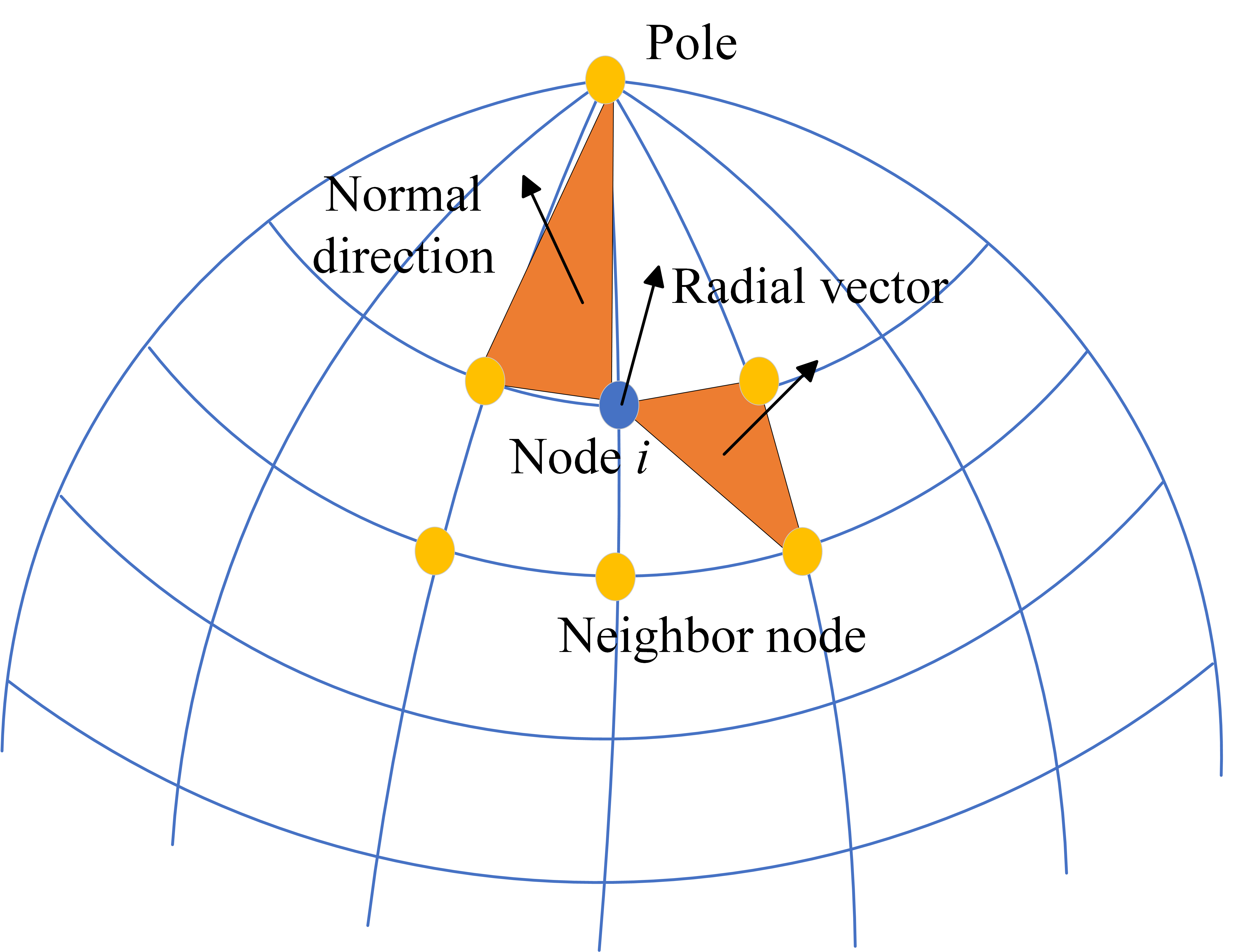}
			\caption{Topological structure and orientation definition for positions on the first latitude circle.}
			\label{fig11}
		\end{minipage}
	\vspace{-0.4cm}
\end{figure*}
To generate a set of discrete positions satisfying the aforementioned constraints on a spherical surface, a Fibonacci sphere-based method was proposed in \cite{ref14}. However, such automatically generated positions lack natural topological relationships, requiring the calculation of pairwise distances to manually construct connections, which complicates the in-depth analysis of antenna movement and time costs. 
To address this issue, we propose a deterministic discretization scheme based on a latitude-longitude grid. Assume the BS is located at the origin, and the 6DMA antenna surfaces are deployed on a sphere of radius $r_0$ centered at the BS. The sphere is divided into $F$ equally spaced meridians, with the set of azimuth angles denoted by $\Phi = \{\phi_f = 2\pi f/F\}_{f=0}^{F-1}$.

To satisfy the minimum inter-antenna spacing constraint $d_{\min}$, we first determine the first valid latitude circle closest to the pole. This latitude must satisfy two conditions simultaneously. First, the chord length between adjacent points on the latitude (with an azimuthal interval of $2\pi/F$) must not be less than $d_{\min}$, leading to the constraint on the cylindrical radius of the latitude circle:
\begin{equation}
	r_{\text{first}} = \frac{d_{\min}}{2\sin(\pi/F)}.
\end{equation}
Second, the chord length from the pole to any point on this latitude must be at least $d_{\min}$. Based on the chord length formula $s = 2r_0 \sin(\theta/2)$, the corresponding polar distance (the polar angle between the pole and the latitude) must satisfy:
\begin{equation}
	2r_0 \sin(\theta_{\text{first}} / 2) \geq d_{\min},
\end{equation}
combining the above two conditions, we set:
\begin{equation}
	\theta_{\text{first}} = \max\left\{ \arcsin\left( \frac{d_{\min}}{2r_0 \sin(\pi/F)} \right), \ 2\arcsin\left( \frac{d_{\min}}{2r_0} \right) \right\}.
\end{equation}
The corresponding axial distance from the pole is given by:
\begin{equation}
	d_{\text{pole}} = r_0 (1 - \cos\theta_{\text{first}}).
\end{equation}
To ensure that the axial distance between the first northern latitude and the first southern latitude is no less than $d_{\min}$, we must verify:
\begin{equation}
	2r_0 \cos\theta_{\text{first}} \geq d_{\min},
\end{equation}
otherwise, the configuration is infeasible. The remaining axial length available for dividing intermediate latitudes is:
\begin{equation}
	D = 2r_0 \cos\theta_{\text{first}}.
\end{equation}
By fixing the axial spacing between adjacent latitude circles to $d_{\min}$, the number of intermediate latitudes is determined as:
\begin{equation}
	L = \left\lfloor \frac{D}{d_{\min}} \right\rfloor,
\end{equation}
where $\lfloor \cdot \rfloor$ denotes the floor function. Since the chord length between two points on a sphere is strictly greater than the difference in their $z$-coordinates, and the $z$-difference here is defined as $d_{\min}$, the distance between any points on adjacent latitude circles exceeds $d_{\min}$, thereby satisfying constraint \eqref{3}.
The total number of latitude circles includes the first northern latitude, the first southern latitude, and the $L$ intermediate latitudes, totaling $L+2$. Including the North and South Poles, the total number of discrete positions is:
\begin{equation}
	M = F \cdot (L+2) + 2.
\end{equation}
Each non-pole position is represented in spherical coordinates as $(r_0, \theta_g, \phi_f)$, where $\theta_g$ denotes the polar angle of the $g$-th latitude circle, with $g=0$ corresponding to the first northern latitude and $g=L+1$ corresponding to the first southern latitude.

As shown in Fig.~\ref{fig9}, to define the neighborhood relationship, an 8-neighbor structure is established for each non-pole position and its adjacent polar positions, covering upper/lower neighbors on the same meridian, left/right neighbors on the same latitude, and four diagonal neighbors. For any current point $\mathbf{q}_i$ and its two neighbors $\mathbf{q}_{i_1}, \mathbf{q}_{i_2}$ forming a triangle, the unit normal vector is calculated as:
\begin{equation}
	\mathbf{n}_i^{(m)} = \frac{(\mathbf{q}_{i_1} - \mathbf{q}_i) \times (\mathbf{q}_{i_2} - \mathbf{q}_i)}{\|(\mathbf{q}_{i_1} - \mathbf{q}_i) \times (\mathbf{q}_{i_2} - \mathbf{q}_i)\|}, \quad m=1,\ldots,8.
\end{equation}
In addition, each position includes a radial outward normal vector:
\begin{equation}
	\mathbf{n}_i^{(0)} = \frac{\mathbf{q}_i}{\|\mathbf{q}_i\|},
\end{equation}
resulting in a total of $J=9$ discrete orientations for each position. Fig.~\ref{fig10} illustrates the neighbor definition for the poles, where all points on the first latitude circle serve as neighbors, yielding $J=F+1$ discrete orientations. Fig.~\ref{fig11} defines the neighbors for the first latitude circle, which includes the pole, left/right neighbors on the same latitude, and three points on the adjacent latitude (on the same and adjacent meridians), resulting in $J=7$ discrete orientations. The generation of discrete orientations follows a similar method to that of intermediate positions, determining rotation directions via triangular facets formed with neighbors, and including the radial direction.

Based on this, we can generate a graph according to the topological relationships between discrete positions and further analyze the movement patterns and costs of the antennas using graph theory. Let the set of discrete positions be $\mathcal{Q} = \{\mathbf{q}_i\}_{i=1}^M$, corresponding to an undirected adjacency graph $\mathcal{G} = (\mathcal{Q}, \mathcal{E})$, where an edge $(i,j) \in \mathcal{E}$ exists if and only if $\mathbf{q}_j \in \mathcal{N}_i$, i.e., position $j$ is a neighbor of position $i$.
\subsection{Definition and Proof of Lower Bounds on Movement Cost and Time Cost}
Given the current antenna configuration set $\mathbf{Z}(t)$ and the next configuration set $\mathbf{Z}(t+1)$, since the antennas are indistinguishable, the transition from $\mathbf{Z}(t)$ to $\mathbf{Z}(t+1)$ can be viewed as an unlabeled Multi-Agent Path Finding (MAPF) problem \cite{mapf, 1207121}. Since the number of discrete positions is typically much larger than the number of configured antenna surfaces, the probability of mutual blocking during movement is low. Thus, the specific movement pattern is not difficult to solve.

To define the movement and time costs of the configuration transition, we assume that the unit energy consumption for a single antenna surface to move to an adjacent discrete position is $\Delta E$, and the unit time duration is $\Delta t_{\text{step}}$. The rotational transformation of the antenna surface can be completed simultaneously with the positional movement and is therefore not considered separately. 
The lower bounds of the movement cost and time cost from configuration $\mathbf{Z}(t)$ to $\mathbf{Z}(t+1)$ can be exactly solved using the BFS algorithm and the Hungarian algorithm.

The shortest path for any antenna surface from an initial position $\mathbf{q}_{i_u}$ belonging to $\mathbf{Z}(t)$ to a target position $\mathbf{q}_{j_v}$ belonging to $\mathbf{Z}(t+1)$ can be determined via BFS on graph $\mathcal{G}$. The path length (in steps) is defined as:
\begin{equation}
	d_{u,v} = \text{BFS}(\mathbf{q}_{i_u}, \mathbf{q}_{j_v}) \in \mathbb{N}.
\end{equation}
By running BFS for every pair of initial and target positions $(i_u, j_v)$, we obtain the distance matrix $\mathbf{D} = [d_{u,v}] \in \mathbb{N}^{U \times U}$.
The total number of steps required for movement can be calculated using the Hungarian algorithm. Due to the high connectivity of the graph, there may exist multiple assignment schemes with the minimum total steps for the transition from $\mathbf{Z}(t)$ to $\mathbf{Z}(t+1)$. To simultaneously ensure the minimum time cost, we slightly modify the Hungarian algorithm.
We define a cost matrix with a penalty term:
\begin{equation}
	C_{u,v} = d_{u,v} + \kappa d_{u,v}^2,
\end{equation}
where $\kappa > 0$ is a penalty coefficient. By incorporating the quadratic term $d_{u,v}^2$, we aim to construct a lexicographical optimization mechanism: the primary objective is to minimize the total movement steps (energy cost), while the secondary objective is to minimize the variance of the movement steps, which in turn minimizes the maximum displacement of any single antenna and thereby minimizes the time cost.
The validity of this cost function is rigorously established through the following propositions.

\textbf{Proposition 1 (Priority of Total Energy Efficiency):} 
To ensure that the algorithm strictly prioritizes the minimization of total movement steps $E_{\min}$ over the penalty term, the coefficient $\kappa$ is required to satisfy the condition:
\begin{equation}
	\kappa < \frac{1}{U \cdot d_{\max}^2},
\end{equation}
where $d_{\max}$ represents the diameter of the graph $\mathcal{G}$ (the maximum possible distance between any two nodes).

\textit{Proof:} Let $\mathcal{X}_A$ and $\mathcal{X}_B$ be two distinct valid matching schemes. Let $S_A = \sum_{(u,v)\in\mathcal{X}_A} d_{u,v}$ and $P_A = \sum_{(u,v)\in\mathcal{X}_A} d_{u,v}^2$ denote the total steps and the penalty sum for scheme A, respectively (similarly for scheme B).
Assume that scheme A is superior in terms of energy efficiency, i.e., $S_A < S_B$. Since the steps are integers, the minimum difference is $S_B \geq S_A + 1$.
To guarantee that the modified Hungarian algorithm selects scheme A, the total cost $J_A$ needs to be strictly less than $J_B$:
\begin{equation}
	S_A + \kappa P_A < S_B + \kappa P_B.
\end{equation}
Considering the worst-case scenario where $S_B = S_A + 1$ and the penalty difference is maximized (i.e., $P_A$ approaches its theoretical maximum $P_{\max}$ while $P_B \to 0$), the condition becomes:
\begin{equation}
	S_A + \kappa P_{\max} < S_A + 1 + 0 \implies \kappa < \frac{1}{P_{\max}}.
\end{equation}
The theoretical maximum penalty occurs when all $U$ antennas move the maximum distance $d_{\max}$, yielding $P_{\max} = U \cdot d_{\max}^2$. Thus, setting $\kappa < (U \cdot d_{\max}^2)^{-1}$ ensures that the weighted penalty term never exceeds the cost of a single movement step, thereby preserving the optimality of the total energy consumption. \hfill $\blacksquare$

\textbf{Proposition 2 (Optimization of Time Cost via Load Balancing):} 
For any two schemes with the same total movement steps (i.e., $S_A = S_B$), the cost function with a quadratic penalty ($n=2$) ensures that the solution with a more balanced step distribution (smaller maximum step) yields a lower total cost.

\textit{Proof:} This property is derived from Majorization Theory. Let the movement step vectors of two schemes be sorted in descending order: $\mathbf{d}_A = (d_1^A, \dots, d_U^A)$ and $\mathbf{d}_B = (d_1^B, \dots, d_U^B)$.
If scheme A has a more uneven distribution (e.g., a larger maximum step) than scheme B while maintaining the same sum, vector $\mathbf{d}_A$ is said to \textit{majorize} $\mathbf{d}_B$, denoted as $\mathbf{d}_A \succ \mathbf{d}_B$.
According to the Hardy-Littlewood-Pólya inequality, for any strictly convex function $f(\cdot)$, if $\mathbf{d}_A \succ \mathbf{d}_B$, then:
\begin{equation}
	\sum_{u=1}^U f(d_u^A) > \sum_{u=1}^U f(d_u^B).
\end{equation}
In our cost function, the penalty term $f(d) = \kappa d^2$ has a second derivative $f''(d) = 2\kappa > 0$, confirming that it is strictly convex. Therefore, the scheme with the more balanced distribution (scheme B) will result in a lower total penalty sum. Consequently, the modified Hungarian algorithm will automatically prefer the matching scheme that suppresses outliers, thereby minimizing the maximum single-antenna movement steps $d_{\max}$ and reducing the time cost $T_{\min}$. \hfill $\blacksquare$

The matching variable is defined as:
\begin{equation}
	x_{u,v} = \begin{cases} 
		1, & \text{if antenna moves from } \mathbf{q}_{i_u} \to \mathbf{q}_{j_v}, \\
		0, & \text{otherwise}.
	\end{cases}
\end{equation}
The optimal matching is obtained by solving the following linear assignment problem:
\begin{equation}
	\begin{aligned}
		\min_{\{x_{u,v}\}} \quad & \sum_{u=1}^U \sum_{v=1}^U C_{u,v} x_{u,v} \\
		\text{s.t.} \quad & \sum_{v=1}^U x_{u,v} = 1, \quad \forall u, \\
		& \sum_{u=1}^U x_{u,v} = 1, \quad \forall v, \\
		& x_{u,v} \in \{0,1\}.
	\end{aligned}
\end{equation}
Using the Hungarian algorithm, the optimal matching $\hat{\mathbf{X}} = [\hat{x}_{u,v}]$ can be found in polynomial time. Let the corresponding distance matrix after matching be:
\begin{equation}
	\mathbf{D}^{\text{assigned}} = \mathbf{D} \odot \hat{\mathbf{X}},
\end{equation}
where $\odot$ denotes the element-wise product, representing the distances of the selected pairs.
The total movement steps (lower bound on energy cost) is defined as:
\begin{equation}
	E_{\min} = \Delta E \sum_{u=1}^U \sum_{v=1}^U \hat{x}_{u,v} d_{u,v},
\end{equation}
which is the sum of movement steps of all antennas multiplied by the unit energy consumption. The maximum single-antenna movement steps can be calculated as:
\begin{equation}
	d_{\max} = \max_{u=1,\ldots,U} \sum_{v=1}^U \hat{x}_{u,v} d_{u,v},
\end{equation}
corresponding to the steps required by the slowest moving antenna. Therefore, the lower bound on time cost is defined as:
\begin{equation}
	T_{\min} = \Delta t_{\text{step}} \times d_{\max}.
\end{equation}
Here, it is assumed that all antennas start moving simultaneously, and the overall time is determined by the longest path.

\begin{algorithm}[t]
	\caption{Adaptive Single-Step Reconfiguration with Historical Feedback}
	\label{alg:adaptive_reconfig}
	\small 
	\SetAlgoLined
	\KwIn{Offline mapping library $\{\mathcal{C}_g\}$, update interval $N$, initial warmup periods $L_{\text{init}}$, history weight factor $\omega$, max simulation time $T_{\max}$.}
	\KwOut{Dynamic antenna configuration $\mathbf{Z}(t)$.}
	
	\textbf{Initialize:} Current configuration $\mathbf{Z} \leftarrow \text{Default}$, History library $\mathcal{H} \leftarrow \emptyset$, Period index $l \leftarrow 0$\;
	
	\For{$t \leftarrow 1$ \KwTo $T_{\max}$}{
		Update positions of all vehicles $\mathbf{p}_k(t)$ based on kinematic model\;
		
		\If{$(t-1) \pmod N == 0$}{
			$l \leftarrow l + 1$\;
			Predict vehicle trajectories for the upcoming period $\mathcal{T}_l$\;
			Identify candidate search space $\bigcup_{u=1}^U \bar{\mathcal{N}}_{\text{pos}_u}$ based on $\mathbf{Z}$ and neighbors\;
			
			\eIf{$l \leq L_{\text{init}}$}{
				Set current weight $\omega_{\text{curr}} \leftarrow 0$ (Use offline prior only)\;
			}{
				Set current weight $\omega_{\text{curr}} \leftarrow \omega$ (Enable historical feedback)\;
			}
			
			\ForEach{position $i \in \bigcup_{u=1}^U \bar{\mathcal{N}}_{\text{pos}_u}$}{
				Calculate offline prediction score $S^{\text{pre}}_i$ based on active grids\;
				Retrieve historical rate $\bar{R}_i$ from $\mathcal{H}$ to compute $S^{\text{hist}}_i$\;
				Compute composite score $S_i$ via Eq.~\eqref{score} using $\omega_{\text{curr}}$\;
			}
			
			Select $U$ positions with highest scores to form $\mathcal{I}_{\text{new}}$\;
			Determine optimal rotation $j^*$ using offline stats\;
			Update antenna configuration $\mathbf{Z} \leftarrow \mathbf{Z}_{\text{new}}$\;
		}
		
		Execute uplink transmission and measure sum rate\;
		
		\If{$t \pmod N == 0$}{
			Calculate average rate $R_l$ for current period\;
			Update history library $\mathcal{H} \leftarrow \mathcal{H} \cup \{(\mathcal{I}_{\text{new}}, R_l)\}$\;
		}
	}
\end{algorithm}
\section{CSI-Free 6DMA Reconfiguration Method} \label{method}

In this section, we propose a fast, CSI-free antenna configuration method. This approach utilizes offline collected data to determine the initial antenna positions and dynamically adjusts the configuration based on both offline data and historical service data to adapt to the mobility characteristics of vehicle users.

\subsection{Offline Environment Profiling and Space-Configuration Mapping Modeling}

Given the directional sparsity characteristic of 6DMA systems in complex scattering environments, users located in a specific physical space can typically achieve high-gain service only through a limited set of position-angle combinations. By exploiting the relatively static nature of macroscopic scatterers like buildings in urban environments, we can pre-construct a static mapping library that links each spatial grid to a preferred antenna configuration, serving as prior knowledge for online decision-making.

First, the ground service area $\Omega = [0, X] \times [0, Y]$ is discretized into uniform grids with a side length of $W$. The total number of grids is denoted by $N_{\text{grid}} = \lceil X/W \rceil \cdot \lceil Y/W \rceil$. For a grid with index $g = (g_x, g_y)$, its geometric center coordinates are set to $\mathbf{c}_{g} = [(g_x - 0.5)W, (g_y - 0.5)W, z_{\text{veh}}]^\top$, where $z_{\text{veh}}$ represents the typical antenna height of a vehicle.
To quantify the configuration gain, $S$ typical user positions $\{\mathbf{p}_g^{(s)}\}_{s=1}^S$ are uniformly sampled within each grid $g$. For each tuple $(\mathbf{q}_i, \mathbf{r}_j)$ in the global discrete configuration set $\mathcal{P}$, the average theoretical spectral efficiency within the grid is calculated as:
\begin{equation}
	\bar{r}_g(i,j) = \frac{1}{S} \sum_{s=1}^S B \log_2 \left( 1 + \Gamma_k \bigr( \mathbf{H}(\{{\mathbf{p}}_g^{(s)}\}_{s=1}^S), (\mathbf{q}_i, \mathbf{r}_j) \bigr) \right),
\end{equation}
where the $\Gamma_k$ is calculated according to \eqref{SINR} based on static large-scale channel parameters. Theoretically, traversing all grids and feasible configurations allows for the establishment of a complete prior map. However, due to the massive number of combinations involving discrete positions $M$ and rotational states $J$, an exhaustive search over the entire domain faces prohibitively high computational complexity.

To address this issue, we propose a geometric pruning and hierarchical search strategy based on the geometric properties of 6DMAs to dynamically constrain the search space:

\begin{enumerate}
	\item Geometric Hemispherical Pruning: Utilizing the LoS relationship between the BS and the target grid, obvious back-facing invalid positions are eliminated. The antenna positions are restricted to the grid-facing hemisphere, constructing a valid position subset $\mathcal{Q}_g^h$:
	\begin{equation}
		\mathcal{Q}_g^h = \left\{ \mathbf{q}_i \in \mathcal{Q} \mid \mathbf{n}_i^\top (\mathbf{c}_g - \mathbf{q}_{\text{BS}}) \geq 0 \right\},
	\end{equation}
	where $\mathbf{q}_{\text{BS}}$ denotes the BS coordinates, and $\mathbf{n}_i$ is the reference normal vector at position $\mathbf{q}_i$. This constraint ensures that the angle between the antenna position and the grid center does not exceed $90^\circ$.
	
	\item Two-Stage Hierarchical Sampling: Based on $\mathcal{Q}_g^h$, the candidate scale is further reduced.
	\begin{itemize}
		\item \textit{Coarse-Grained Screening}: Using the K-means clustering algorithm, $Y$ representative anchor positions are uniformly selected within $\mathcal{Q}_g^h$. The average rates of these positions under the default radial orientation are calculated to quickly evaluate their coverage potential, from which the top $N_{\text{seed}}$ seed positions with the highest rates are screened out.
		\item \textit{Fine-Grained Expansion}: Centered on these $N_{\text{seed}}$ seed positions, an expanded search is conducted within their first-order physical neighborhood (which is still required to satisfy the hemispherical constraint) to discover local optimal positions.
	\end{itemize}
	
	\item Rotation Refinement: For the high-potential positions identified through the two-stage screening, all feasible rotational states $j \in \{1, \dots, J\}$ are traversed to finally determine the optimal position-rotation combination.
\end{enumerate}

Through the above three-level search strategy, the directional sparsity of the 6DMA is fully utilized. While ensuring that the candidate set covers high-performance configurations, the search space is compressed by several orders of magnitude. Finally, for each grid $g$, the top $H$ configurations with the highest rates are retained to form the offline preferred candidate set:
\begin{equation}
	\mathcal{C}_g = \left\{ (\mathbf{q}_i, \mathbf{r}_j) \in \mathcal{P} \mid \bar{r}_g(i,j) \text{ ranks Top-} H \right\}.
\end{equation}
This prior library $\mathcal{C}_g$ accurately captures the directional response characteristics under stable scattering environments and macroscopic geometry, providing a high-quality and compact decision-making basis for subsequent online antenna scheduling.

\subsection{Online Adaptive Optimization Based on Historical Rates}

Although the offline mapping library provides optimal solution suggestions for static environments, instantaneous channel perturbations caused by vehicle movement and prediction errors may lead to deviations between actual performance and prior estimates. To enhance the system robustness in dynamic environments, this paper proposes an adaptive configuration scheme fusing offline prior and online feedback. This scheme relies not only on the current predicted distribution but also makes full use of historical periodic measured data. By accumulatively learning from historical configuration performance, the evaluation of channel quality for each candidate position is continuously corrected, thereby guiding the antenna position and rotation decisions for the $(l)$-th period.

\subsubsection{Multi-Dimensional Hybrid Scoring Mechanism}
Recalling the problem formulation in Section~\ref{question}, the feasible search space for the $u$-th antenna is restricted to its closed neighborhood $\bar{\mathcal{N}}_{\text{pos}_u}$. To determine the optimal configuration, we first construct a comprehensive scoring function $S_i$ for every unique candidate position $i$ contained within the union of these neighborhoods (i.e., $i \in \bigcup_{u=1}^U \bar{\mathcal{N}}_{\text{pos}_u}$). This score is a weighted sum of the offline prediction score $S^{\text{pre}}_i$, the historical accumulated measured score $S^{\text{hist}}_i$, and the stability reward $B^{\text{stab}}_i$:
\begin{equation} \label{score}
	S_i = (1-\omega) S^{\text{pre}}_i + \omega S^{\text{hist}}_i + B^{\text{stab}}_i,
\end{equation}
where $\omega \in [0,1]$ is a balancing factor used to adjust the weights of the offline prior and online historical feedback.

\paragraph{Offline Prediction Score}
This term aims to index the offline library using the predicted vehicle distribution. Let $\mathcal{G}_{\text{active}}$ denote the set of active grids covered by the predicted trajectory for the $(l)$-th period, where the predicted demand density of grid $g$ is $\rho_g$ and the maximum density is $\rho_{\max}$. We evaluate the hit status of position $i$ in the preferred sets $\mathcal{C}_g$ of these grids:
\begin{equation}
	S^{\text{pre}}_i = \sum_{g \in \mathcal{G}_{\text{active}}} \frac{\rho_g}{\rho_{\max}} \cdot \left[ \beta_0 + \beta_1 (\nu_{g,i} - 1) \right] \cdot \mathbb{I}(i \in \mathcal{C}_g),
\end{equation}
where $\mathbb{I}(\cdot)$ is the indicator function, indicating whether position $i$ exists in $\mathcal{C}_g$, and $\nu_{g,i}$ represents the number of distinct rotational states associated with position $i$ in $\mathcal{C}_g$. Here, $\beta_1 > 0$ is a multi-modal reward coefficient, which tends to favor robust positions that perform excellently under multiple rotations.

\paragraph{Historical Accumulated Rate Score}
This term utilizes historical data to correct prior biases. A constantly growing historical record library $\mathcal{H}_{l-1} = \{(\mathcal{I}_{k}, R_{k}) \mid k = 1, \dots, l-1\}$ is maintained, where $R_{k}$ is the measured system average sum rate in the $k$-th period.
For any candidate position $i$, its score depends on the average performance when activated in all past periods. The historical average contribution rate $\bar{R}_i$ is calculated as:
\begin{equation}
	\bar{R}_i = 
	\begin{cases} 
		\frac{\sum_{k=1}^{l-1} \mathbb{I}(i \in \mathcal{I}_{k}) R_{k}}{\sum_{k=1}^{l-1} \mathbb{I}(i \in \mathcal{I}_{k})}, & \text{if } \sum_{k=1}^{l-1} \mathbb{I}(i \in \mathcal{I}_{k}) > 0, \\
		\bar{R}_{\text{global}}, & \text{otherwise},
	\end{cases}
\end{equation}
where the denominator counts the total number of times position $i$ has been selected in history, and the numerator is the corresponding accumulated rate. If position $i$ has never been selected (i.e., a cold-start position), it is filled with the current global historical average rate $\bar{R}_{\text{global}} = \frac{1}{l-1}\sum_{k=1}^{l-1} R_k$ to provide an unbiased initial estimate.
The normalized historical score is given by $S^{\text{hist}}_i = \bar{R}_i / R_{\text{norm}}$, where $R_{\text{norm}}$ is the historical peak rate.

\paragraph{Stability Reward}
To suppress frequent mechanical switching caused by marginal performance gains (the ping-pong effect), an inertia reward is granted to positions activated in the current period:
\begin{equation}
	B^{\text{stab}}_i = 
	\begin{cases} 
		\mu, & \text{if } i \in \mathcal{I}_{l-1}, \\
		0, & \text{otherwise},
	\end{cases}
\end{equation}
where $\mu$ is the stability threshold.

\subsubsection{Greedy Position Assignment and Decoupled Rotation Decision}
Based on the comprehensive score $S_i$, we employ a position-rotation decoupling strategy to generate the new configuration $\mathbf{Z}_{l}$ and approximate the optimal solution with minimal computational complexity. We first perform position selection using a conflict-aware greedy algorithm that respects the individual movement constraints. Specifically, for each antenna surface $u$ ($u=1,\dots,U$) activated at position $\text{pos}_u$ during the previous $(l-1)$-th period, we traverse all candidate positions $i$ belonging to its specific closed neighborhood $\bar{\mathcal{N}}_{\text{pos}_u}$. We then identify the candidate $i^*$ that yields the maximum score $S_i$ within this local subset (i.e., $i^* = \arg\max_{i \in \bar{\mathcal{N}}_{\text{pos}_u}} S_i$) as the intended target for antenna $u$. To resolve potential conflicts where neighborhoods overlap, we prioritize the antenna-position assignments with higher scores to form the final non-conflicting position set $\mathcal{I}_{l}$.

Following this, the rotation decision involves tallying the occurrence of rotation states in the currently active grids for each position $i \in \mathcal{I}_{l}$ via the offline library:
\begin{equation}
	m_{i,j} = \sum_{g \in \mathcal{G}_{\text{active}}} \mathbb{I}\left( (\mathbf{q}_i, \mathbf{r}_j) \in \mathcal{C}_g \right).
\end{equation}
The rotation direction $j^* = \arg\max_{j} m_{i,j}$ with the highest accumulated frequency is selected as the optimal solution. If no prior match exists where $m_{i,j}=0$, the system defaults to the radial direction pointing toward the BS, specifically the $j$ that minimizes $\mathbf{n}_j^\top \mathbf{q}_i$. We then finalize the update by setting $[\mathbf{Z}_{l}]_{i,j^*} = 1$.

The detailed procedure is outlined in \AlgRef{alg:adaptive_reconfig}. At the initialization stage, offline mapping solely determines the initial antenna positions. The quality of this initial solution is of critical importance to the proposed single-step 6DMA system. Upon accumulating sufficient online data, the weighting factor for historical feedback is restored to its nominal level. Additionally, the antenna reconfiguration interval $N$ defines the prediction horizon for the cumulative vehicle distribution.
\begin{table}[htbp]
	\footnotesize
	\caption{Values of parameters}
	\label{tab1}
	\centering
	\begin{tabular}{|c|c|c|c|}
		\hline
		\textbf{Parameter} & \textbf{Value} & \textbf{Parameter} & \textbf{Value} \\
		\hline
		$K$ & 30/35/40/45/50 & $v$ & $10$--$20\,\mathrm{m/s}$ \\
		\hline
		$U$ & $16$ & $Q$ & $4\,(2\times2)$ \\
		\hline
		$B$ & $20\,\mathrm{MHz}$ & $N$ & $20/10/1$ \\
		\hline
		$d_{min}$ & $0.1\,\mathrm{m}$ & $r_0$ & $0.5\,\mathrm{m}$ \\
		\hline
		$X \times Y$ & $300\,\mathrm{m} \times 300\,\mathrm{m}$ & $W$ & $15\,\mathrm{m}$ \\
		\hline
		$F$ & $12$ & $S$ & $20$ \\
		\hline
		$\phi_{3\mathrm{dB}}$ & $65^\circ$ & $\theta_{3\mathrm{dB}}$ & $25^\circ$ \\
		\hline
		$G_{\max}$ & $8.0\,\mathrm{dBi}$ & & \\
		\hline
	\end{tabular}
\end{table}

\section{Simulation Results} \label{results}

In this section, we evaluate the performance of the proposed 6DMA system in IoV scenarios through numerical simulations implemented in Python 3.8. We consider a typical urban intersection scenario where roadside buildings serve as the primary reflectors. It is worth emphasizing that while the antenna configuration is optimized based on the predicted vehicle trajectories, the simulation performance metrics are evaluated using the actual vehicle positions. This approach validates the robustness of the proposed scheme against prediction errors. Detailed simulation parameters are summarized in Table~\ref{tab1}.

\begin{figure}[t]
	\centering
	\vspace{-0.1cm}
	\includegraphics[width=0.8\columnwidth]{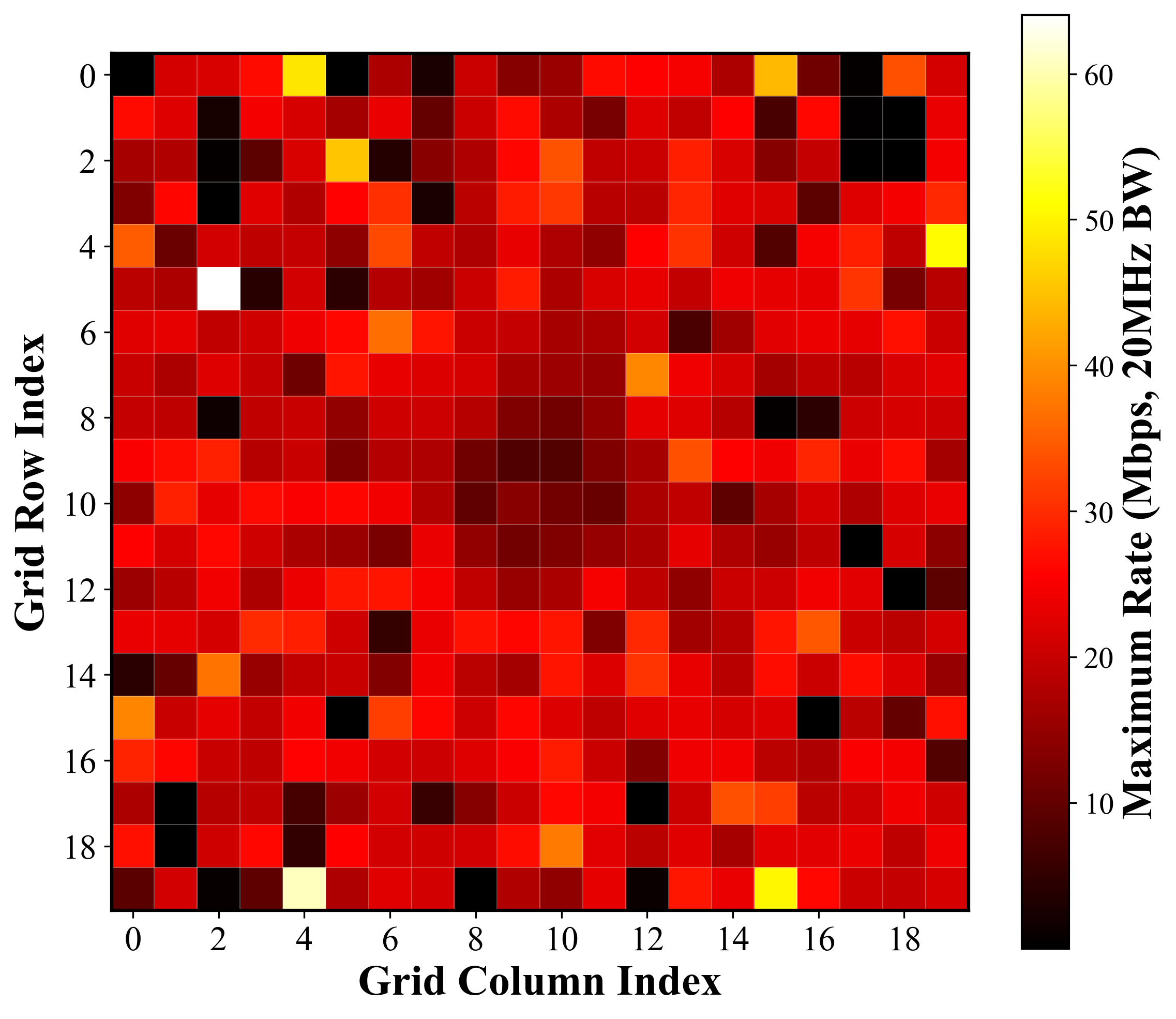}
	\caption{Grids Performance Heatmap}
	\label{fig2}
\end{figure}
To validate the superiority of the proposed method, we compare the following five schemes:
\begin{enumerate}
	\item Fixed Position Antenna: A static deployment scheme serving as a benchmark. This scheme consists of four $90^\circ$ sectors, each equipped with a fixed rectangular antenna array and a uniform $15^\circ$ downtilt. The total number of antenna elements is the same as that of the 6DMA system.
	\item Circular Position 6DMA: Four sectorized surfaces move along a fixed circular track parallel to the ground, with a downtilt of $15^\circ$.
	\item Discrete-Rotation-Only 6DMA: The positions of the four surfaces are fixed, but their orientations can be adjusted at discrete intervals within a predefined angular range.
	\item Full Reconfiguration 6DMA: A global search baseline scheme. This scheme includes 16 movable surfaces (each being an array with the same total number of elements as the FPA). At each reconfiguration instance, the next position for every surface is searched from the entire discrete space.
	\item Proposed Single-Step 6DMA: The scheme proposed in this paper. Its hardware parameters are identical to the Full Reconfiguration scheme, but at each update, the search is restricted to the current position and its one-hop neighborhood, as defined in \eqref{eq:prob_local}.
\end{enumerate}

Fig.~\ref{fig2} illustrates the mapping from the user grid to the antenna configuration, established based on directional sparsity. Each square represents a physical location in the scenario, and the shade of the color represents the average rate achievable by a single antenna surface that provides the best coverage for that square. Users located directly beneath the BS suffer from relatively low rates due to excessive incident angles and lower antenna gains. The annular region surrounding the BS yields the highest rates due to low path loss and favorable incident angles. Conversely, cell-edge users exhibit the poorest performance due to severe path loss and the lack of LoS links.

\begin{figure}[t]
	\centering
	\vspace{-0.1cm}
	\includegraphics[width=0.9\columnwidth]{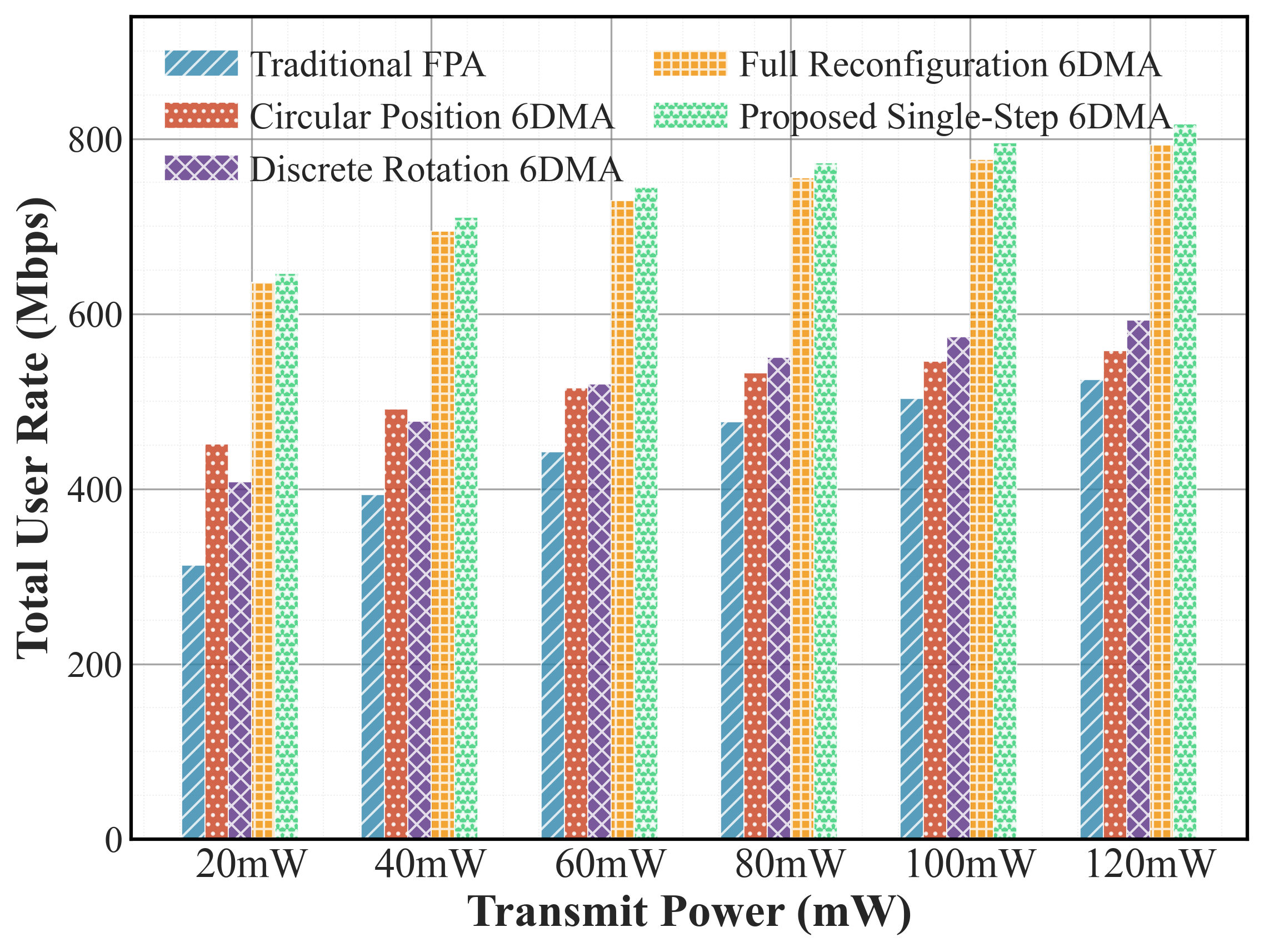}
	\caption{Comparison of uplink sum rates versus transmit power for different antenna schemes.}
	\label{fig4}
\end{figure}

Fig.~\ref{fig4} presents the variation of the system sum rate with transmit power for a scenario with 30 vehicles. The proposed 6DMA scheme significantly outperforms the traditional FPA, indicating that the high spatial flexibility of the 6DMA can effectively adapt to user distributions and capture large-scale channel variations. Although the Circular-Track and Discrete-Rotation-Only 6DMA schemes perform better than the FPA, their gains are constrained by limited spatial degrees of freedom. It is noteworthy that the optimization method based on single-step movement achieves a higher rate than the full reconfiguration method. This is attributed to the fact that the single-step method has a smaller action space, making it easier to find a near-optimal solution, whereas the full reconfiguration method involves a vast action space where heuristic approaches often fail to fully exploit its potential.

\begin{figure}[t]
	\centering
	\vspace{-0.1cm}
	\includegraphics[width=0.9\columnwidth]{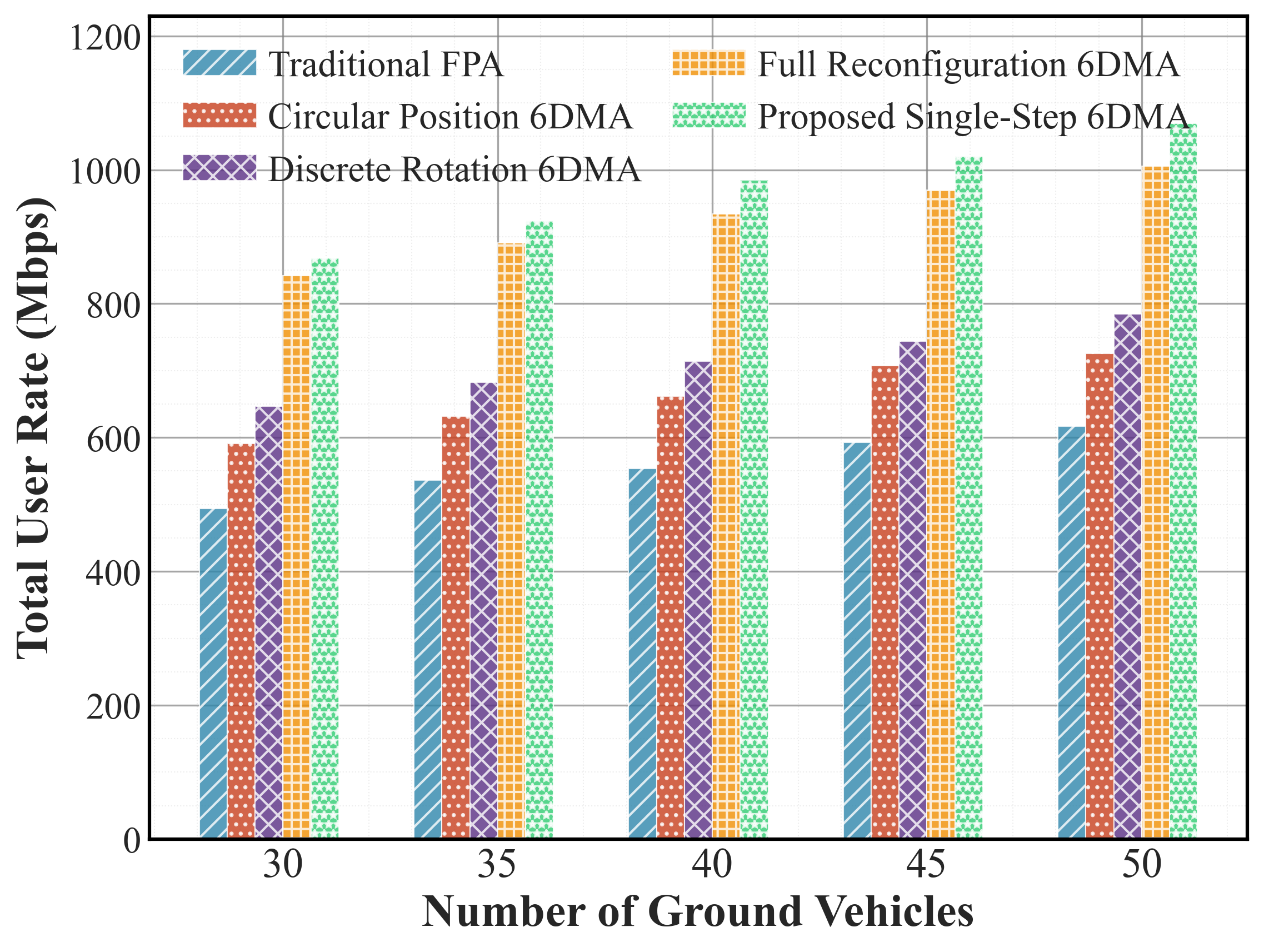}
	\caption{System sum rate versus number of vehicle users under fixed transmit power.}
	\label{fig3}
\end{figure}
Fig.~\ref{fig3} illustrates the trend of the system sum rate with respect to the number of users under a fixed transmit power. As the number of vehicles increases, the sum rate rises rapidly but eventually saturates due to intensified inter-user interference. The proposed 6DMA framework consistently demonstrates the best performance among all baselines. This is because vehicle distributions in road environments are typically dispersed. The 6DMA system can leverage its spatial flexibility to dynamically concentrate antenna resources on relatively clustered areas, whereas physically constrained baseline schemes fail to provide such targeted services.

\begin{figure}[t]
	\centering
	\vspace{-0.1cm}
	\includegraphics[width=0.9\columnwidth]{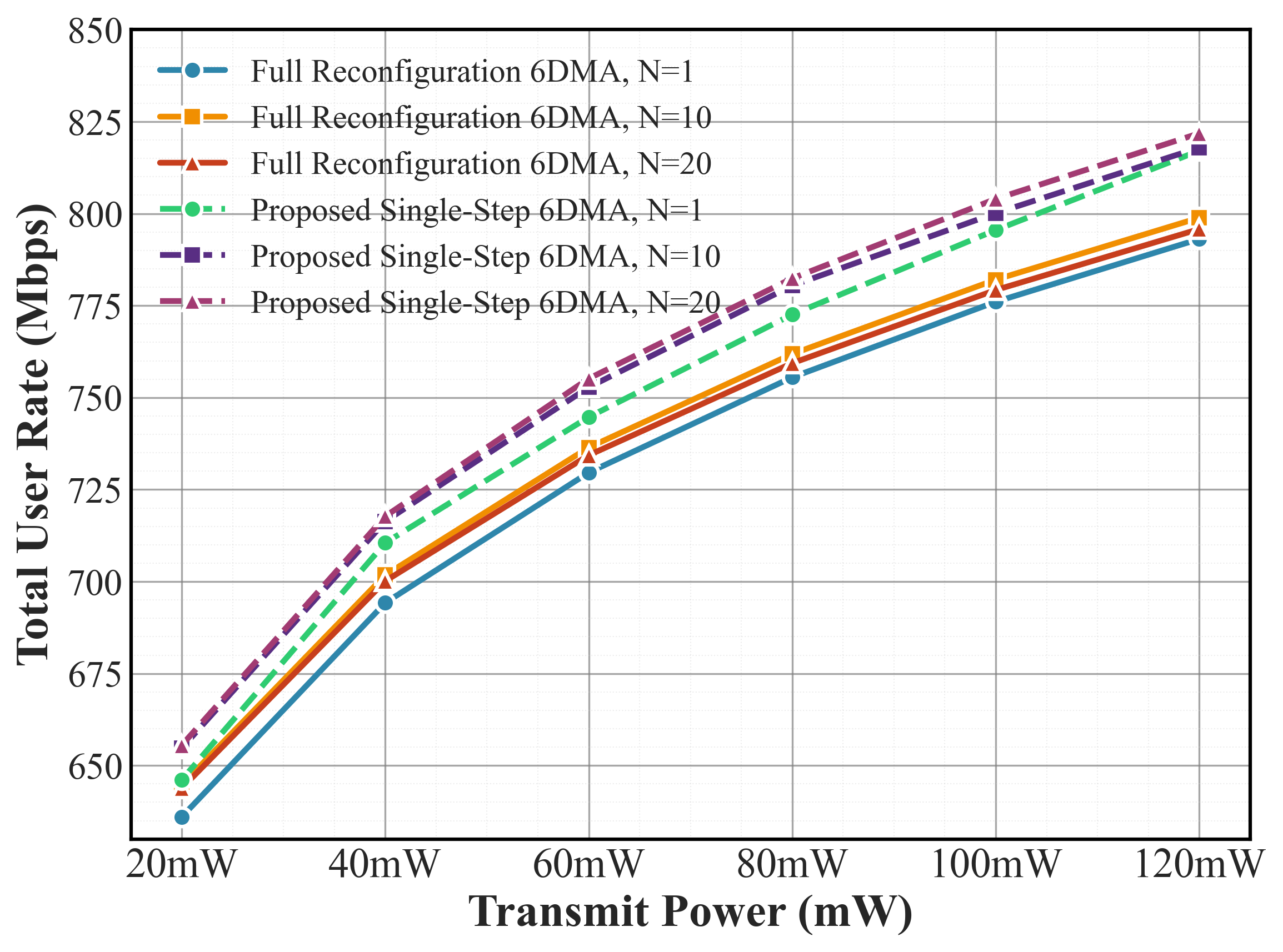}
	\caption{Impact of reconfiguration interval $N$ on sum rate under different transmit powers.}
	\label{fig5}
\end{figure}

Fig.~\ref{fig5} analyzes the system performance under different transmit powers with a fixed number of 30 vehicle users, varying the antenna update interval $N$. It can be observed that the proposed single-step movement scheme comprehensively outperforms the heuristic global reconfiguration scheme. Notably, the achievable rate at $N=1$ is lower than that at $N=10$ and $N=20$. This is because, at $N=1$, the optimization relies on the instantaneous sparse distribution of a few vehicles, resulting in a low-density grid map where it is difficult for the algorithm to discriminate between superior antenna positions. In contrast, for $N=10$ and $20$, the input is the cumulative predicted distribution over future time slots. This aggregation provides a statistically more robust heatmap of user hotspots, offering a better basis for decision-making even in the presence of minor prediction errors. This further demonstrates the feasibility of antenna configuration based on predicted distributions, which not only reduces the reconfiguration frequency but also enhances performance.

\begin{figure}[t]
	\centering
	\vspace{-0.1cm}
	\includegraphics[width=0.9\columnwidth]{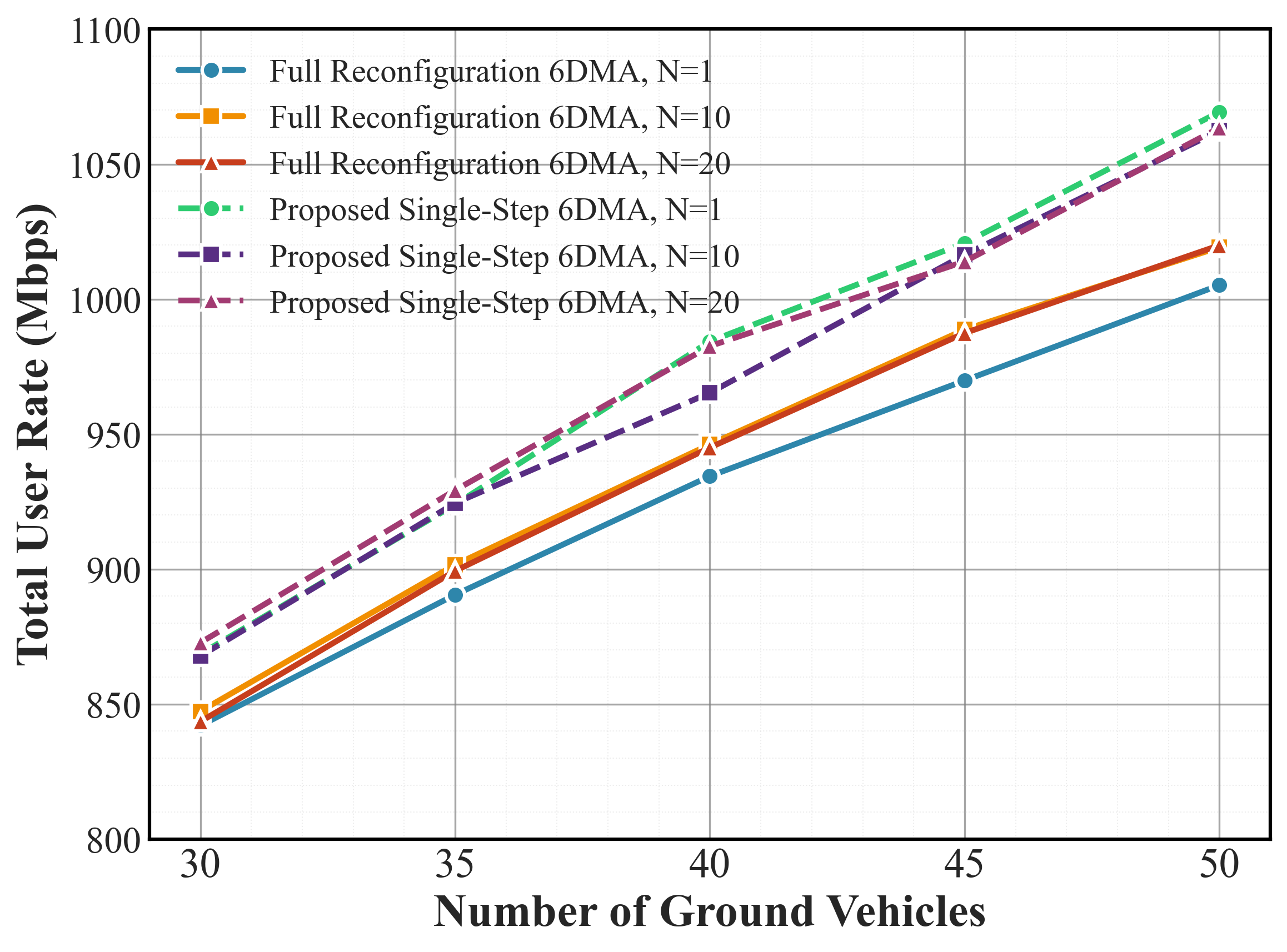}
	\caption{Impact of reconfiguration interval $N$ on sum rate under different vehicle densities.}
	\label{fig6}
\end{figure}
\begin{figure}[t]
	\centering
	\vspace{-0.1cm}
	\includegraphics[width=0.9\columnwidth]{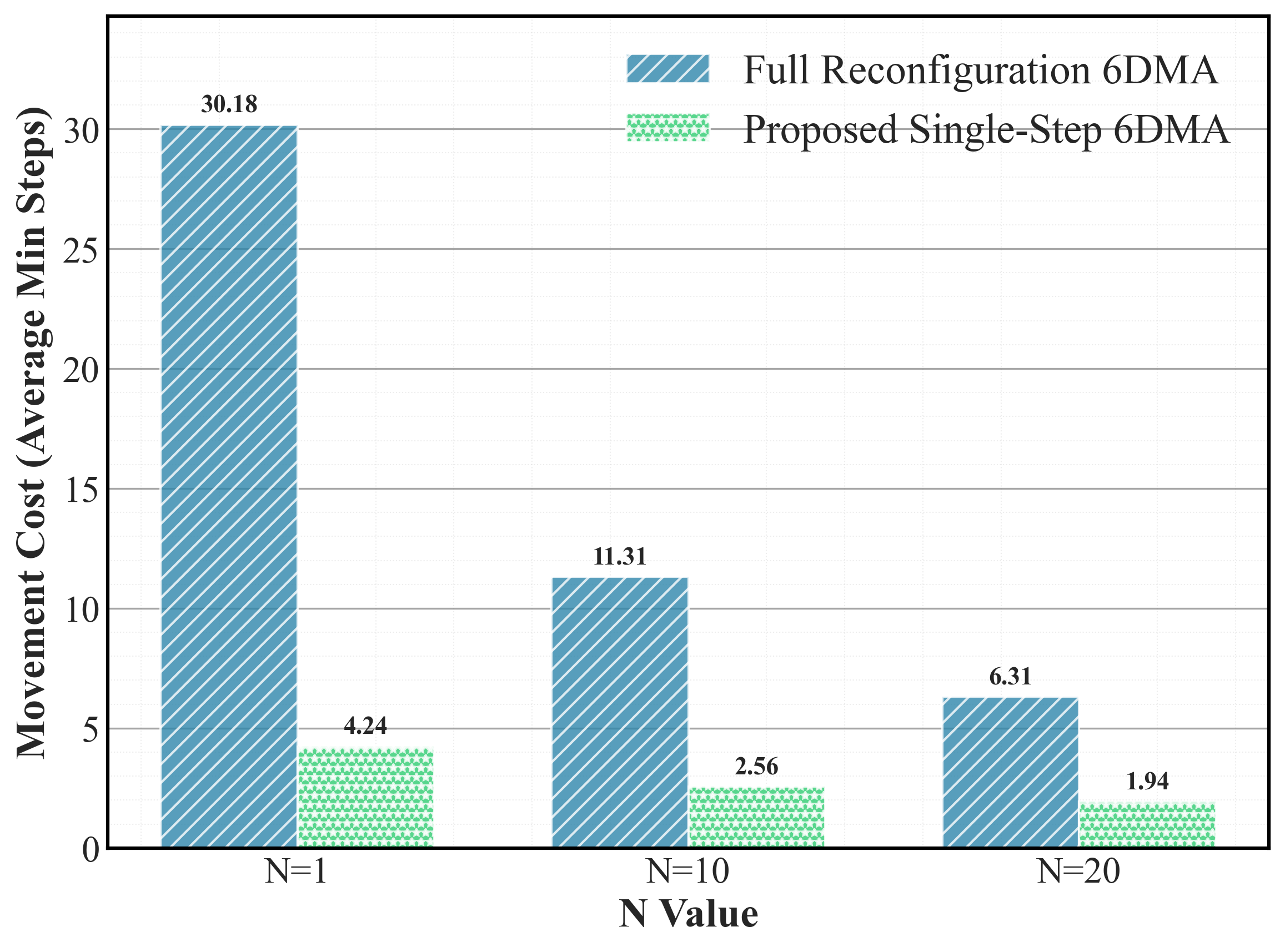}
	\caption{Average movement cost versus reconfiguration interval $N$.}
	\label{fig7}
\end{figure}

\begin{figure}[t]
	\centering
	\vspace{-0.1cm}
	\includegraphics[width=0.9\columnwidth]{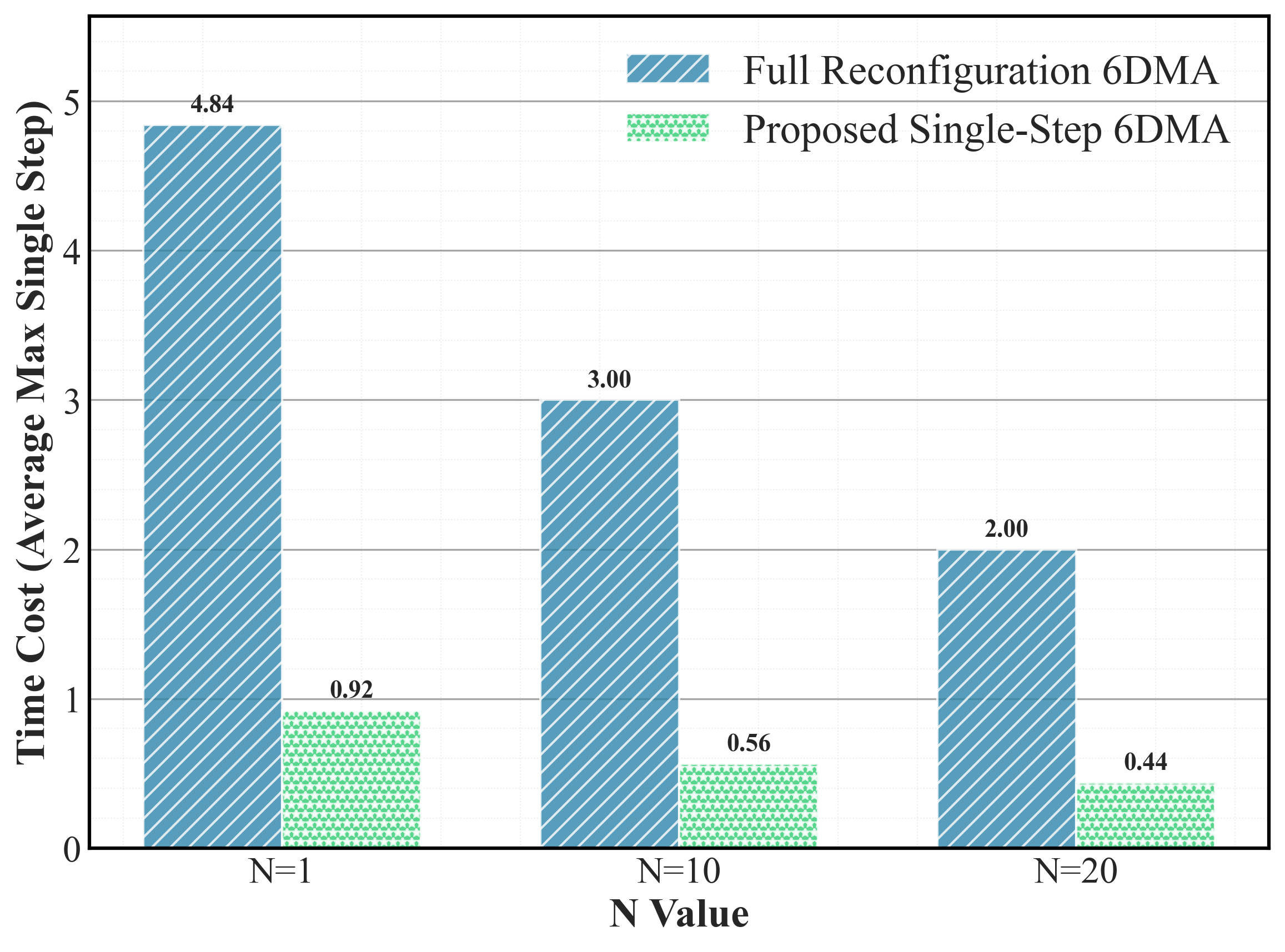}
	\caption{Average time cost versus reconfiguration interval $N$.}
	\label{fig8}
\end{figure}

Fig.~\ref{fig6} analyzes the system performance under different vehicle user counts with a fixed transmit power of 23 dBm, varying the antenna update interval $N$. The experiment employed 10 random seeds, with fifty simulations conducted for each seed, and the global mean was taken. The results indicate that while the value of $N$ has some impact on system performance, it is not significant, and a higher $N$ value is slightly superior to $N=1$ when the number of users is small. As user density increases, a smaller reconfiguration interval begins to show a slight advantage, as it can better track fast-changing microscopic dynamics. However, from a practical implementation perspective, a larger $N$ can significantly reduce physical overhead with negligible performance loss. The single-step method consistently outperforms the full reconfiguration scheme, demonstrating excellent local adaptability.

Fig.~\ref{fig7} analyzes the theoretical lower bound of the average movement cost required for each position update for both the single-step movement method and the full reconfiguration method. According to the modeling in Section~\ref{sec:discrete_generation}, moving each antenna to a neighbor position incurs one unit of cost. When $N$ is low, the adjustment of the antenna configuration is based on the instantaneous actual vehicle user distribution. Due to the high mobility of vehicles, the grids containing users may change significantly between adjacent decision intervals, thus requiring substantial adjustments to antenna positions, leading to high movement costs and frequent reconfiguration. As $N$ increases, the cumulative user distribution exhibits stable macroscopic patterns, resulting in minimal distribution changes between adjacent decision moments, and consequently, lower antenna movement amplitudes. The average movement cost of the antenna configuration method based on single-step movement is in the single digits, implying that adapting to a new distribution requires moving only a few antennas. This provides strong theoretical support for the practical deployment of 6DMA.

Fig.~\ref{fig8} analyzes the theoretical lower bound of the time cost required for each antenna reconfiguration for the two methods. The time cost is defined as the number of steps required by the antenna with the longest moving path during reconfiguration. Similar to the movement cost, the time cost shows a significant downward trend as $N$ increases. The minimum value for the time cost of a single reconfiguration should be $1$. However, values less than $1$ appearing in the single-step movement method imply that for some reconfigurations, no changes in antenna positions occurred. This reflects that the variation in the predicted cumulative vehicle user distribution is limited, as the macroscopic distribution remains approximately static within the prediction window. By capturing long-term patterns rather than chasing instantaneous fluctuations, the system avoids unnecessary mechanical adjustments, demonstrating high implementability.

\section{Conclusion} \label{conclusion}

In this paper, we introduced 6DMA antenna technology into IoV to improve the achievable rate. Based on graph theory, we modeled the movement cost and latency cost of 6DMA for the first time. To address the challenges of fast-varying channels, we proposed a low-complexity, CSI-free optimization framework. This framework leverages predicted vehicle distribution information to enable rapid antenna deployment by restricting the search space to a single-step spatial neighborhood. Simulation results demonstrate that the proposed scheme significantly outperforms traditional baseline schemes. Furthermore, the analysis indicates that the prediction-based method typically requires moving only a small number of antennas, or even none, during reconfiguration, thereby demonstrating its robustness, low mechanical overhead, and good practical feasibility in highly dynamic wireless networks.



\begin{thebibliography}{00}
	
	\bibitem{ref1}
	H. Li, K. Ota, and M. Dong, ``Learning IoV in 6G: Intelligent edge computing for Internet of Vehicles in 6G wireless communications,'' \emph{IEEE Wireless Commun.}, vol. 30, no. 6, pp. 96--101, Dec. 2023.
	
	\bibitem{001}
	Q.~Wu, H.~Liu, C.~Zhang, Q.~Fan, Z.~Li, and K.~Wang, ``Trajectory protection schemes based on a gravity mobility model in IoT,'' \emph{Electronics}, vol. 8, no. 2, p. 148, 2019.
	
	\bibitem{002}
	Q.~Wu and J.~Zheng, ``Performance modeling and analysis of the ADHOC MAC protocol for VANETs,'' in \emph{Proc. IEEE Int. Conf. Commun. (ICC)}, London, UK, 2015, pp. 3646--3652.
	
	\bibitem{003}
	Q.~Wu, S.~Xia, Q.~Fan, and Z.~Li, ``Performance analysis of IEEE 802.11 p for continuous backoff freezing in IoV,'' \emph{Electronics}, vol. 8, no. 12, p. 1404, 2019.
	
	\bibitem{004}
	Q.~Wu, S.~Nie, P.~Fan, H.~Liu, F.~Qiang, and Z.~Li, ``A swarming approach to optimize the one-hop delay in smart driving inter-platoon communications,'' \emph{Sensors}, vol. 18, no. 10, p. 3307, 2018.
	
	\bibitem{005}
	Q.~Wu and J.~Zheng, ``Performance modeling and analysis of the ADHOC MAC protocol for vehicular networks,'' \emph{Wireless Netw.}, vol. 22, no. 3, pp. 799--812, 2016.
	
	\bibitem{006}
	Q.~Wu and J.~Zheng, ``Performance modeling and analysis of IEEE 802.11 DCF based fair channel access for vehicle-to-roadside communication in a non-saturated state,'' \emph{Wireless Netw.}, vol. 21, no. 1, pp. 1--11, 2015.
	
	\bibitem{007}
	Q.~Wu and J.~Zheng, ``Performance modeling of the IEEE 802.11 p EDCA mechanism for VANET,'' in \emph{Proc. IEEE Global Commun. Conf. (GLOBECOM)}, Austin, TX, USA, 2014, pp. 57--63.
	
	\bibitem{008}
	Z.~Zhang, Q.~Wu, P.~Fan, N.~Cheng, W.~Chen, and K.~B.~Letaief, ``DRL-based optimization for AoI and energy consumption in C-V2X enabled IoV,'' \emph{IEEE Trans. Green Commun. Netw.}, early access, 2025.
	
	
	\bibitem{ref2}
	X. Chen, Y. Deng, H. Ding, G. Qu, H. Zhang, P. Li, and Y. Fang, ``Vehicle as a Service (VaaS): Leverage vehicles to build service networks and capabilities for smart cities,'' \emph{IEEE Commun. Surveys Tuts.}, vol. 26, no. 3, pp. 2048--2081, 3rd Quart., 2024.
	
	\bibitem{ref3}
	C. Li, M. Dong, Y. Fu, F. R. Yu, and N. Cheng, ``Integrated sensing, communication, and computation for IoV: Challenges and opportunities,'' \emph{IEEE Commun. Surveys Tuts.}, early access, 2025, doi: \href{https://doi.org/10.1109/COMST.2025.3612388}{10.1109/COMST.2025.3612388}.
	
		\bibitem{ref013}
	Q.~Wang, D.~O.~Wu, and P.~Fan, ``Delay-constrained optimal link scheduling in wireless sensor networks,'' \emph{IEEE Trans. Veh. Technol.}, vol. 59, no. 9, pp. 4564--4577, 2010.
	\bibitem{ref014}
	W.~Li, J.~Li, and P.~Fan, ``Network coding for two-way relaying networks over Rayleigh fading channels,'' \emph{IEEE Trans. Veh. Technol.}, vol. 59, no. 9, pp. 4476--4488, 2010.
	\bibitem{ref015}
	J.~Zhang, P.~Fan, and K.~B.~Letaief, ``Network coding for efficient multicast routing in wireless ad-hoc networks,'' \emph{IEEE Trans. Commun.}, vol. 56, no. 4, pp. 598--607, 2008.
	\bibitem{ref016}
	Z.~Yao, J.~Jiang, P.~Fan, Z.~Cao, and V.~O.~K.~Li, ``A neighbor-table-based multipath routing in ad hoc networks,'' in \emph{Proc. 57th IEEE Semiannu. Veh. Technol. Conf. (VTC Spring)}, Jeju, South Korea, 2003, pp. 1739--1743.
	\bibitem{ref017}
	P.~Fan, C.~Feng, Y.~Wang, and N.~Ge, ``Investigation of the time-offset-based QoS support with optical burst switching in WDM networks,'' in \emph{Proc. IEEE Int. Conf. Commun. (ICC)}, New York, NY, USA, 2002, pp. 2682--2686.
	\bibitem{ref018}
	P.~Fan and X.-G.~Xia, ``Block coded modulation for the reduction of the peak to average power ratio in OFDM systems,'' in \emph{Proc. IEEE Wireless Commun. Netw. Conf. (WCNC)}, New Orleans, LA, USA, 1999, pp. 1095--1099.
	\bibitem{ref019}
	Q.~Wu, X.~Wang, Q.~Fan, P.~Fan, C.~Zhang, and Z.~Li, ``High stable and accurate vehicle selection scheme based on federated edge learning in vehicular networks,'' \emph{China Commun.}, vol. 20, no. 3, pp. 1--17, 2023.
	\bibitem{ref020}
	X.~Di, K.~Xiong, P.~Fan, H.~C.~Yang, and K.~B.~Letaief, ``Optimal resource allocation in wireless powered communication networks with user cooperation,'' \emph{IEEE Trans. Wireless Commun.}, vol. 16, no. 12, pp. 7936--7949, 2017.
	
	\bibitem{10233705} 
	C. Meng, K. Xiong, W. Chen, B. Gao, P. Fan, and K. B. Letaief, ``Sum-rate maximization in STAR-RIS-assisted RSMA networks: A PPO-based algorithm,'' \emph{IEEE Internet Things J.}, vol. 11, no. 4, pp. 5667--5680, 2024. 
	
	\bibitem{5720555}
	H. Zhou, P. Fan, and J. Li, ``Global proportional fair scheduling for networks with multiple base stations,'' \emph{IEEE Trans. Veh. Technol.}, vol. 60, no. 4, pp. 1867--1879, 2011.
	
	\bibitem{ref4}
	J.-H. Jo, J.-N. Shim, B. Kim, C.-B. Chae, and D. K. Kim, ``AoA-based position and orientation estimation using lens MIMO in cooperative vehicle-to-vehicle systems,'' \emph{IEEE J. Sel. Areas Commun.}, vol. 41, no. 12, pp. 3719--3735, Dec. 2023.
	
	\bibitem{ref5}
	H. Zhu, Q. Wu, X.-J. Wu, Q. Fan, P. Fan, and J. Wang, ``Decentralized power allocation for MIMO-NOMA vehicular edge computing based on deep reinforcement learning,'' \emph{IEEE Internet Things J.}, vol. 9, no. 14, pp. 12770--12782, Jul. 2022.
	
	
	\bibitem{ref6}
	H. Jiang, Z. Zhang, J. Dang, and L. Wu, ``A novel 3-D massive MIMO channel model for vehicle-to-vehicle communication environments,'' \emph{IEEE Trans. Commun.}, vol. 66, no. 1, pp. 79--90, Jan. 2018.
	
	\bibitem{ref7}
	X. Yu, L. Tu, Q. Yang, M. Yu, Z. Xiao, and Y. Zhu, ``Hybrid beamforming in mmWave massive MIMO for IoV with dual-functional radar communication,'' \emph{IEEE Trans. Veh. Technol.}, vol. 72, no. 7, pp. 9017--9030, Jul. 2023.
	
	\bibitem{ref8}
	D. L\'{o}pez-P\'{e}rez, A. De Domenico, N. Piovesan, G. Xinli, H. Bao, S. Qitao, and M. Debbah, ``A survey on 5G radio access network energy efficiency: Massive MIMO, lean carrier design, sleep modes, and machine learning,'' \emph{IEEE Commun. Surveys Tuts.}, vol. 24, no. 1, pp. 653--697, 1st Quart., 2022.
	
	\bibitem{ref009}
	J.~Fan, S.~Yin, Q.~Wu, and F.~Gao, ``Study on refined deployment of wireless mesh sensor network,'' in \emph{Proc. 6th Int. Conf. Wireless Commun. Netw. Mobile Comput. (WiCOM)}, Chengdu, China, 2010, pp. 1--5.
	
	\bibitem{ref010}
	K.~Xiong, P.~Fan, Z.~Xu, H.~C.~Yang, and K.~B.~Letaief, ``Optimal cooperative beamforming design for MIMO decode-and-forward relay channels,'' \emph{IEEE Trans. Signal Process.}, vol. 62, no. 6, pp. 1476--1489, 2014.
	
	\bibitem{ref011}
	Y.~Yang and P.~Fan, ``Doppler frequency offset estimation and diversity reception scheme of high-speed railway with multiple antennas on separated carriage,'' \emph{J. Mod. Transport.}, vol. 20, no. 4, pp. 227--233, 2012.
	
	\bibitem{ref012}
	H.~Zhou, P.~Fan, and J.~Li, ``Global proportional fair scheduling for networks with multiple base stations,'' \emph{IEEE Trans. Veh. Technol.}, vol. 60, no. 4, pp. 1867--1879, 2011.
	
	\bibitem{ref99}
	L. Zhu, W. Ma, W. Mei, Y. Zeng, Q. Wu, B. Ning, Z. Xiao, X. Shao, J. Zhang, and R. Zhang, ``A tutorial on movable antennas for wireless networks,'' \emph{IEEE Commun. Surveys Tuts.}, early access, 2025, doi: 
	\href{https://doi.org/10.1109/COMST.2025.3546373}{10.1109/COMST.2025.3546373}.
	
	\bibitem{ref9}
	Y. Gao, Q. Wu, W. Mei, G. Chen, W. Chen, and Z. Zheng, ``Integrating movable antennas and intelligent reflecting surfaces for coverage enhancement,'' \emph{IEEE Trans. Wireless Commun.}, Early Access, 2025, doi: 
	\href{https://doi.org/10.1109/TWC.2025.3623474}{10.1109/TWC.2025.3623474}.
	
	\bibitem{Li2025Joint}
	Z. Li, J. Ba, Z. Su, H. Peng, Y. Wang, W. Chen, and Q. Wu, ``Joint discrete antenna positioning and beamforming optimization in movable antenna enabled full-duplex ISAC networks,'' \emph{IEEE Trans. Wireless Commun.}, Early Access, 2025, doi:
	\href{https://doi.org/10.1109/TWC.2025.3630154}{10.1109/TWC.2025.3630154}. 
	
	\bibitem{zhu2025multiuser}
	L. Zhu, H. Sun, W. Ma, Z. Xiao, and R. Zhang, ``Multiuser communications aided by cross-linked movable antenna array: Architecture and optimization,'' \emph{IEEE Trans. Wireless Commun.}, early access, 2025, doi:
	\href{https://doi.org/10.1109/TWC.2025.3626388}{10.1109/TWC.2025.3626388}.
	
	\bibitem{ref10}
	K. Wong and K. Tong, ``Fluid antenna multiple access,'' \emph{IEEE Trans. Wireless Commun.}, vol. 21, no. 7, pp. 4801--4815, Jul. 2022.
	
	\bibitem{ref11}
	L. Zhu, W. Ma, and R. Zhang, ``Movable-antenna array enhanced beam forming: Achieving full array gain with null steering,'' \emph{IEEE Commun. Lett.}, vol. 27, no. 12, pp. 3340--3344, Dec. 2023.
	
	\bibitem{li2025over}
	N. Li, P. Wu, B. Ning, L. Zhu, and W. Mei, ``Over-the-air computation via 2-D movable antenna array,'' \emph{IEEE Wireless Commun. Lett.}, vol. 14, no. 1, pp. 33--37, Jan. 2025.
	
	\bibitem{ref12}
	X. Shao, Q. Jiang, and R. Zhang, ``6D movable antenna based on user distribution: Modeling and optimization,'' \emph{IEEE Trans. Wireless Commun.}, vol. 24, no. 1, pp. 355--370, Jan. 2025.
	
	\bibitem{shao20256dma}
	X. Shao and R. Zhang, ``6DMA enhanced wireless network with flexible antenna position and rotation: Opportunities and challenges,'' \emph{IEEE Commun. Mag.}, vol. 63, no. 4, pp. 121--128, Apr. 2025.
	
	\bibitem{ref13}
	X. Shao, W. Mei, C. You, Q. Wu, B. Zheng, C.-X. Wang, J. Li, R. Zhang, R. Schober, L. Zhu, W. Zhuang, and X. Shen, ``A tutorial on six-dimensional movable antenna for 6G networks: Synergizing positionable and rotatable antennas,'' \emph{IEEE Commun. Surveys Tuts.}, early access, 2025, doi: 
	\href{https://doi.org/10.1109/COMST.2025.3602939}{10.1109/COMST.2025.3602939}.
	
	\bibitem{ref14}
	X. Shao, R. Zhang, Q. Jiang, and R. Schober, ``6D movable antenna enhanced wireless network via discrete position and rotation optimization,'' \emph{IEEE J. Sel. Areas Commun.}, vol. 43, no. 3, pp. 674--687, Mar. 2025.
	
	\bibitem{ref15}
	X. Shao, R. Zhang, Q. Jiang, J. Park, T. Q. S. Quek, and R. Schober, ``Distributed channel estimation and optimization for 6D movable antenna: Unveiling directional sparsity,'' \emph{IEEE J. Sel. Topics Signal Process.}, vol. 19, no. 2, pp. 349--365, Mar. 2025.
	
	\bibitem{ref19}
	H. Hua, Y. Zhou, W. Mei, J. Xu, and R. Zhang, ``Hierarchically tunable 6DMA for wireless communication and sensing: Modeling and performance optimization,'' \emph{IEEE Trans. Wireless Commun.}, early access, 2025, doi: 
	\href{https://doi.org/10.1109/TWC.2025.3613548}{10.1109/TWC.2025.3613548}.
	
	\bibitem{ref16}
	S. Jang and C. Lee, ``New view of learning-aided channel estimation for movable antenna systems,'' \emph{IEEE Trans. Wireless Commun.}, vol. 24, no. 7, pp. 5694--5708, Jul. 2025.
	
	\bibitem{ref17}
	C. Liu, W. Mei, Z. Chen, J. Fang, and B. Ning, ``A general optimization framework for movable antenna systems via discrete sampling,'' \emph{IEEE Wireless Commun. Lett.}, early access, 2025, doi: \href{https://doi.org/10.1109/LWC.2025.3629978}{10.1109/LWC.2025.3629978}.
	
	\bibitem{ref18}
	Q. Jiang, X. Shao, and R. Zhang, ``Low-complexity 6DMA rotation and position optimization based on statistical channel information,'' in \emph{Proc. IEEE/CIC Int. Conf. Commun. China (ICCC Workshops)}, Shanghai, China, 2025, pp. 1--6.
	
	\bibitem{ref20}
	X. Shao, L. Hu, Y. Sun, X. Li, Y. Zhang, J. Ding, X. Shi, F. Chen, D. W. K. Ng, and R. Schober, ``Hybrid near-far field 6D movable antenna design exploiting directional sparsity and deep learning,'' \emph{IEEE Trans. Wireless Commun.}, early access, 2025, doi: \href{https://doi.org/10.1109/TWC.2025.3605550}{10.1109/TWC.2025.3605550}.
	
	\bibitem{ref21}
	3GPP, ``Study on channel model for frequencies from 0.5 to 100 GHz,'' 3rd Generation Partnership Project (3GPP), Tech. Rep. TR 38.901 V14.1.1, Release 14, Aug. 2017. [Online]. Available: \url{http://www.3gpp.org/DynaReport/38901.htm}
	
	\bibitem{mapf}
	J. Lee and W. Chung, ``Challenges in applying multi-agent path finding solutions to real-world applications,'' in \emph{Proc. 24th Int. Conf. Control, Autom. Syst. (ICCAS)}, Jeju, South Korea, 2024, pp. 1160--1161.
	
	\bibitem{1207121} 
	Z. Yao, J. Jiang, P. Fan, Z. Cao, and V. O. K. Li, ``A neighbor-table-based multipath routing in ad hoc networks,'' in \emph{Proc. IEEE 57th Semiannual Veh. Technol. Conf. (VTC Spring)}, Jeju, South Korea, 2003, vol. 3, pp. 1739--1743. 
\end{thebibliography}
\end{document}